\documentclass[letterpaper]{article} % DO NOT CHANGE THIS
\usepackage{aaai25}  % DO NOT CHANGE THIS
\usepackage{times}  % DO NOT CHANGE THIS
\usepackage{helvet}  % DO NOT CHANGE THIS
\usepackage{courier}  % DO NOT CHANGE THIS
\usepackage[hyphens]{url}  % DO NOT CHANGE THIS
\usepackage{graphicx} % DO NOT CHANGE THIS
\usepackage{natbib}  % DO NOT CHANGE THIS AND DO NOT ADD ANY OPTIONS TO IT
\usepackage{caption} % DO NOT CHANGE THIS AND DO NOT ADD ANY OPTIONS TO IT
\usepackage{algorithm}
\usepackage{algorithmic}
\usepackage{newfloat}
\usepackage{listings}
\DeclareCaptionStyle{ruled}{labelfont=normalfont,labelsep=colon,strut=off} % DO NOT CHANGE THIS
\floatstyle{ruled}
\newfloat{listing}{tb}{lst}{}
\floatname{listing}{Listing}
\title{How May U.S. Courts Scrutinize Their Recidivism Risk Assessment Tools? \\ Contextualizing AI Fairness Criteria on a Judicial Scrutiny-based Framework}
\author {
    Tin Nguyen\textsuperscript{\rm 1},
    Jiannan Xu\textsuperscript{\rm 2},
    Phuong-Anh Nguyen-Le\textsuperscript{\rm 3},\\
    Jonathan Lazar\textsuperscript{\rm 3},
    Donald Braman\textsuperscript{\rm 4},
    Hal Daumé III\textsuperscript{\rm 1},
    Zubin Jelveh\textsuperscript{\rm 3, \rm 5}
}
\affiliations {
    \textsuperscript{\rm 1}Department of Computer Science, University of Maryland, College Park, Maryland, USA\\
    \textsuperscript{\rm 2}Robert H. Smith School of Business, University of Maryland, College Park, Maryland, USA\\
    \textsuperscript{\rm 3}College of Information, University of Maryland, College Park, Maryland, USA\\
    \textsuperscript{\rm 4} George Washington University Law School, Washington DC, USA\\
    \textsuperscript{\rm 5}Department of Criminology and Criminal Justice, University of Maryland, College Park, Maryland, USA\\
    tintn@umd.edu, jiannan@umd.edu, nlpa@umd.edu, \\jlazar@umd.edu, dbraman@law.gwu.edu, hal3@umd.edu, zjelveh@umd.edu
}
\usepackage{booktabs} 
\usepackage[final]{pdfpages}
\usepackage{amsmath}
\usepackage{xspace}
\usepackage{xcolor}
\usepackage{float}

\begin{document}

\maketitle
% \author{AAAI Press\\
% Association for the Advancement of Artificial Intelligence\\
% 2275 East Bayshore Road, Suite 160\\
% Palo Alto, California 94303\\
% }

% \author{Tin Nguyen}
% \email{tintn@umd.edu}
% \affiliation{%
%   \institution{University of Maryland}
%   \city{College Park}
%   \state{Maryland}
%   \country{USA}
% }

% \author{Jiannan Xu}
% \email{jiannan@umd.edu}
% \affiliation{%
%   \institution{University of Maryland}
%   \city{College Park}
%   \state{Maryland}
%   \country{USA}
% }

% \author{Phuong-Anh Nguyen-Le}
% \email{nlpa@umd.edu}
% \affiliation{%
%   \institution{University of Maryland}
%   \city{College Park}
%   \state{Maryland}
%   \country{USA}
% }

% \author{Jonathan Lazar}
% \email{jlazar@umd.edu}
% \affiliation{%
%   \institution{University of Maryland}
%   \city{College Park}
%   \state{Maryland}
%   \country{USA}
% }

% \author{Donald Braman}
% \email{dbraman@law.gwu.edu}
% \affiliation{%
%   \institution{George Washington University Law School}
%   \city{Washington}
%   \state{DC}
%   \country{USA}
% }

% \author{Hal Daumé III}
% \email{hal3@umd.edu}
% \affiliation{%
%   \institution{University of Maryland}
%   \city{College Park}
%   \state{Maryland}
%   \country{USA}
% }

% \author{Zubin Jelveh}
% \email{zjelveh@umd.edu}
% \affiliation{%
%   \institution{University of Maryland}
%   \city{College Park}
%   \state{Maryland}
%   \country{USA}
% }

\begin{abstract}
The AI/HCI and legal communities have developed largely independent conceptualizations of fairness. This conceptual difference hinders the potential incorporation of technical fairness criteria (e.g., procedural, group, and individual fairness) into sustainable policies and designs, particularly for high-stakes applications like recidivism risk assessment. To foster common ground, we conduct legal research to identify if and how technical AI conceptualizations of fairness surface in primary legal sources. We find that while major technical fairness criteria can be linked to constitutional mandates such as ``Due Process'' and ``Equal Protection'' thanks to judicial interpretation, several challenges arise when operationalizing them into concrete statutes/regulations. 
These policies often adopt procedural and group fairness but ignore the major technical criterion of individual fairness. 
Regarding procedural fairness, judicial ``scrutiny'' categories are relevant but may not fully capture how courts scrutinize the use of demographic features in potentially discriminatory government tools like RRA.
Furthermore, some policies contradict each other on whether to apply procedural fairness to certain demographic features. 
Thus, we propose a new framework, integrating U.S. demographics-related legal scrutiny concepts and technical fairness criteria, and contextualize it in three other major AI-adopting jurisdictions (EU, China, and India). 
\end{abstract}

\section{Introduction}

\label{sec:intro}

% \hal{i'm surprised this is diving straight in to RRA... the title and abstract make it sound much more general?}
% \kem{for this paper, I think starting with a tool would be helpful in positioning it as an HCI work, because we say we better the design of these tools by having fairness standings. Abstract updated to be less general}
% \tin{maybe give easier-to-understand examples of RRA and other AI (e.g. credit card fraud detection) tools? per Lazar's comment}

% When AI-assisted decision-making is employed by a government entity instead of 

Recidivism risk assessment (RRA) tools, widely deployed models that predict an individual's likelihood to re-offend following a criminal charge, raise ethical concerns surrounding fairness, transparency, and potential bias. 
These tools rely on quantifiable factors (e.g., criminal history, age at first offense) to produce a risk score that assists court officials in bail, sentencing, parole, and other correctional decisions. 

Outcomes of RRA systems have a direct impact on the lives of people accused or convicted of crime and an indirect impact on their families and communities: manifesting in the forms of emotional and economic hardship when incarcerated family members serve longer sentences or parole conditions, and in the forms of diminished social capital, disintegration, and cycles of crime and poverty for communities with high rates of incarceration. 

RRAs have traditionally been studied in several fields, namely Psychology \cite{hanson2009accuracy}, Criminology \cite{caudy2013well}, Law \cite{nishi2019privatizing}, Statistics \cite{imrey2015commentary}, and other interdisciplinary communities \cite{mann2010assessing, sreenivasan2000actuarial, hamilton2016designed}. 
%The HCI community can contribute meaningfully to this discourse and tackle the examination of fairness in RRA because we are uniquely positioned to address the sociotechnical nature of the problem. Furthermore, in HCI, fairness is a design imperative: HCI research leads in embedding human values into system design, which is crucial for RRA tools. 
In this cross-disciplinary work, spanning Law, AI and Human-Computer Interaction (HCI), we contribute new legal insights for fairness audit of RRA tools.

%introduce to the HCI community theoretical lenses and frameworks, informed by law and fair machine learning, through which researchers in values-centered design can empirically study fairness metrics in RRA systems, further enhance their evaluative powers, and give evidence to inform better designs. 

%Fairness in RRA was conceptualized as “the guarantees [people] [should ask] for as protection against potential bias” when they are “concerned about how [their] decision procedure might operate differentially between two groups of interest” by \citet{kleinberg2017inherent}. 

Research on the fairness of RRA tools has grown significantly across many disciplines since 2016, when an investigative article\footnote{\url{https://www.propublica.org/article/machine-bias-risk-assessments-in-criminal-sentencing}} from a non-profit organization, ProPublica, drew public attention to racial bias in COMPAS, a proprietary AI-based RRA tool used in Florida and many other U.S. states \cite{angwin2022machine}.

Academic research on fairness gained much traction from multiple angles following that publicized project, sometimes by directly criticizing the use of AI for RRA on fairness grounds \cite{dressel2018accuracy, green2019disparate}, other times by subtly pointing out nuanced limitations of fairness criteria (e.g., it is almost impossible to simultaneously optimize for several fairness criteria and/or accuracy \cite{chouldechova2017fair, berk2021fairness}). 

The technical AI literature includes \textbf{procedural fairness} (``fairness through unawareness'' or excluding a feature from model input \cite{bart2024perceptions}) and two mathematical categories of outcome-based fairness: group and individual fairness. Aligned with the intuitive definition by \citet{kleinberg2017inherent}, \textbf{group fairness}, or group parity, is achieved when a statistical metric of interest, e.g., positive outcome rate, is equalized across different groups with respect to a sensitive feature, e.g. race or sex/gender \cite{pedreshi2008discrimination}. \citet{dwork2012fairness} introduced \textbf{individual fairness} based on the intuition that similar individuals should get similar outcomes. Formally, for a pair of individuals, given an input-space distance metric to measure how differently situated they are and an output-space distance metric to measure how different the distributions of their possible outcomes are, individual fairness requires that their output-space distance should be upper-bound by their input-space distance. The more similarly situated two persons are (e.g., same demographics and criminal histories), the more likely that they receive similar outcomes (e.g., RRA scores).
%\hal{say something of similar technical depth to the equalized rates thing for GF} 

%\tin{put these descriptions into related works, or condense into one-sentence summary of these 3 works}
% \citet{chouldechova2017fair} demonstrated that certain group fairness criteria cannot be satisfied at the same time when one group has a higher recidivism prevalence than another.
% \citet{dressel2018accuracy} showed that laypersons may make fairer and more accurate recidivism predictions than COMPAS. 
% \citet{berk2021fairness}  statistically analyzed the relationship between RRA fairness and accuracy and found that accuracy and fairness almost can never be simultaneously maximized. 
%\hal{Overall, these studies showed that...}

%\hal{i think this is wherey ou need to put the "contributions" part.}

%\subsection{Why is this an HCI problem?}

%\hal{is this para really needed?}

%\subsection{Why should HCI address this problem now?} 
%\hal{i'm not really convinced that this subsection belongs in the intro}
%\tin{shifted to Discussion (Policy Implications)?}

%\paragraph{Why U.S. law?}

%\subsection{Contributions}
RRA tools have already been widely adopted across U.S. government bodies for decades. Motivated by the emerging need to identify a constitutional basis for establishing enforceable fairness standards on such AI tools, our main research questions (RQs) are:

\textbf{RQ1.a}: Do technical AI fairness criteria for RRA have an enforceable legal basis under the U.S. Constitution?

\textbf{RQ1.b}: Are there any legal challenges when such abstract fairness criteria are translated into concrete policies?

\textbf{RQ2}: How to develop a new legal framework to mitigate those challenges, informed by established legal theories?

%\textbf{RQ3}: How to design an experiment to validate that framework?

%amid a judicial landscape dominated by the legal interpretation method of ``judicial textualism'' (i.e., narrow interpretation of constitutional/legislative languages, based on plain texts rather than the broad intent of the authors) \cite{scalia2012reading}), 
Two main contributions follow. 
First, we analyzed primary sources of law, including the U.S. Constitution, case law interpreting the Constitution, as well as statutes and regulations at the state and federal levels. The goal is to explore a legal basis for technical fairness criteria, as well as challenges when operationalizing legally relevant concepts of fairness. 
Second, to address the challenges, we propose an integrated framework that combines technical fairness criteria and the judicial ``scrutiny'' concept (typically applied when the government treats people differently based on their demographics). This ``scrutiny'' concept is particularly suited to the RRA use case (compared to other high-stakes AI-assisted decision-making use cases such as loan approval \cite{goyal2024impact} or job application screening \cite{rigotti2024fairness}, which are typically undertaken by private-sector actors) because the U.S. judiciary has developed elaborate scrutiny standards to scrutinize discriminations by public-sector actors (e.g., state and federal government bodies) against private citizens. 
Our framework utilizes a common unit of analysis, scrutiny-worthiness, to help auditors decide on which fairness-related criteria and for which demographics they should prioritize their audit.

\section{Background: Conceptualization of Fairness}

\paragraph{HCI vs. AI}
For a high-level overview, \citet{narayanan2024fairness} conducted a comprehensive review of empirical research on the perceived fairness of AI-assisted decision-making in the organizational setting (e.g., employee performance review, hiring). Their work builds upon the theoretical framework of organizational justice by \citet{colquitt2001dimensionality}, which includes four dimensions: procedural fairness, interpersonal fairness, informational fairness (i.e., explainability), and distributive (i.e., outcome-based) fairness. However, only two of their four fairness dimensions (distributive fairness, and to a lesser extent, procedural fairness) are frequently discussed within the fairness literature of the AI community, as comprehensively reviewed by \citet{pessach2022review}. 
Furthermore, \citet{ryan2023integrating} interviewed roughly equal numbers of AI and HCI experts using procedural fairness and (four mathematical variations of) distributive fairness. They found an interesting theme among the HCI experts' (but not the AI experts) responses: ``Meeting a mathematical definition of fairness does not mean the model is ethical''. This finding suggests that, when humans interact with an AI prediction system to make decisions, i.e., HCI is involved, fairness research should extend beyond technical formulation to consider the social and legal background where the system is deployed. 
% These examples illustrate key differences in how the HCI and AI communities approach the fairness problem of AI-assisted decision-making.

\paragraph{HCI/AI vs. Law}
Legal and technical AI/HCI research on conceptualizations of fairness have evolved mostly on their own, evidenced in the communities having studied similar or neighboring phenomena yet invented different terminologies and frameworks. 
% Some identified challenges in this area of work, arising from the lack of interdisciplinary lenses, include whether existing datasets are suited to serve as a benchmark to standardize the progress of the field and whether a benchmark dataset in an intrinsically sociotechnical context like criminal justice, if it exists, will ensure beneficial and ethical use \cite{bao2021s}. 
% While legal and social science literature recognize the process v. outcome fairness dichotomy \cite{gilliland2002justice}, technical AI/HCI literature contrasts two technical sub-categories of outcome fairness: group (outcome) fairness and individual (outcome) fairness \cite{binns2020apparent}. 
%Group fairness or group parity is achieved when a statistical metric of interest, e.g. positive outcome rate, is equalized across different groups with respect to a sensitive feature, e.g. race or sex \cite{pedreshi2008discrimination}. As a more mathematically involved and therefore harder-to-quantify metric, individual fairness is first introduced by \citet{dwork2012fairness} as a technical concept based on the intuition that similar individuals should be treated with similar outcomes.
For example, a well-received U.S. law review article on RRA fairness maps ``disparate impact'' to outcome fairness, and ``disparate treatment'' both to process (i.e., procedural) fairness and loosely to individual fairness \cite{mayson2019bias}. This mapping implies that individual fairness is not a subcategory of outcome fairness and that individual fairness is more related to procedural fairness than to group fairness. Both are inconsistent with the technical literature. Another seminal work is the Legal chapter (``Understanding United States anti-discrimination law'') in the Fair Machine Learning book by \citet{barocas2023fairness}. This book chapter, however, focuses its investigation of fairness concepts from the perspective of law review articles, and not primary sources of U.S. legal authority (e.g., case law). This is a methodology gap we seek to fill. While \citet{yang2020equal} combined thorough law review and statistical approaches, their work interprets the ``Equal Protection'' clause mostly with procedural fairness and may ignore outcome fairness criteria.%, especially individual fairness. 

Our work contributes to the growing literature on reconciling law and algorithmic fairness. For example, \citet{grossman2024reconciling} evaluated different statistical approaches to assess disparate impact. In addition, \citet{xiang2019legal} demonstrated examples of misalignment between ML fairness concepts and legal definitions.

\section{Methodology -- Legal Research}

In order for technical AI fairness concepts to be adopted into actual RRA audit policies and survive judicial review, we conduct legal research to map these concepts to constitutional concepts.

Legal research can be analogized as a use case of the ``document analysis'' method by \citet{bowen2009document}, which is widely recognized in HCI. Document analysis includes four main steps: 1. ``finding'' documents, 2. ``selecting'' documents, 3. ``appraising'' (making sense of) documents, and 4. ``synthesizing'' data contained in documents. For legal research, ``documents'' include primary sources of U.S. federal and state law, e.g., the U.S. Constitution, case law (judicial branch), statutes (legislative branch), or regulations (executive branch) \cite{barkan2009fundamentals}. 
\citet{linos2017qualitative} laid out common legal research steps (case sampling, case selection, case analysis, and developing theoretical explanation of the case outcomes), which are analogous to the four main steps (documents ``finding'', ``selecting'', ``appraising'', and ``synthesizing'') identified by \citet{bowen2009document}. We apply these four steps to identify relevant court cases and other primary sources of law.

\paragraph{Legal Research in HCI}

Legal research, or legal document analysis, was employed in several HCI studies to inform design practices. \citet{comber2023regulating} analyzed proceedings of a legal dispute in Sweden to study government vs. individual responsibility tension over waste management due to new technology, thereby informing the design of environmental responsibility frameworks among different stakeholders. \citet{gray2021dark} synthesized several sources of European laws (the GDPR, the ePrivacy Directive, and the Court of Justice of the EU) to define what constitutes a ``valid consent'', which serves as a key concept for their subsequent investigation on how ``dark patterns'' (i.e., implicit design to nudge users into selecting a certain privacy setting) in online consent banners should be regulated. \citet{delgado2020sociotechnical} analyzed judicial commentary (e.g., court opinions and law review articles) to understand how Technology-Assisted Review (TAR) during the discovery phase was perceived in the U.S. civil litigation community. \citet{singh2021seeing} adopted the document analysis method by \citet{bowen2009document} on Indian legal documents to study how the Indian government developed Aadhaar, a biometrics-based identification project, to standardize their citizens’ data and deliver government services.

\paragraph{How?}
% AI tools are being used in judicial systems across the world, e.g., China, the U.S., the U.K., the EU, and Russia. However, the U.S. is the only jurisdiction where recidivism risk assessment receives focus \cite{laptev2024application}, and to the best of our knowledge, is popularized at many courts across the country for the purpose of assisting judges in pre-trial bail determination. Therefore, we focus on how primary U.S. legal sources (e.g., the U.S. Constitution, judicial case law, legislative statutes, and executive regulations) conceptualize ``fairness'' in general and in the context of RRA tools in particular. 
For the first step (``finding'' documents), we used a popular U.S. legal research database and search engine, Westlaw, to find primary legal sources with keyword search (such as “fairness”, “Equal Protection”, or “AI”), filters (case law, statutes, regulations), and legal citations in each case found from Westlaw to find more relevant past cases and observe how U.S. courts' positions on the same legal question evolves over time. 
% \tin{case does not include statutes/regulations???}
For the second step (``selecting'' documents), we read the Westlaw-provided summary for each case law or skim through the first few paragraphs of each constitutional provision/statute/regulation to decide if they are actually relevant for our research questions. To ensure that the documents we selected still have legal effects, we also removed any sources of law that have been overturned by later appellate court decisions (marked with a red flag on Westlaw). 
For the third step (``appraising'' documents), we wrote a summary for each selected legal document with relevant excerpts, clarified its legal authority (e.g., binding over which jurisdictions), and substantiated how it is related to fairness, recidivism risk assessment, or AI-assisted decision-making in general (deliverables from this step are in the Appendix). 
For the final step (``synthesizing data''), we discussed our theoretical findings in the next four sections. 
%Once we have come up with the main theoretical findings, we might add some additional case law to further substantiate these established findings.

\paragraph{Interaction among U.S. sources of law}
As explained by \citet{phillips2019practical}, in the U.S., there are four main sources of law at the federal level: the U.S. Constitution (the most binding authority), statutes (from the legislative branch), regulations (from the executive branch) which are more detailed implementation of statutes, and case law (from the judicial branch which follows a hierarchy in order of increasing bindingness: district courts, circuit courts of appeal, and the court of last resort or the Supreme Court) which interpret the Constitution and determine if any statutes or regulations are unconstitutional. Regarding case law, one important nuance is that a decision from a higher court is not binding for a lower court unless the lower court is within the (often geography-defined) jurisdiction of the higher court. For example, case law by the Second Circuit Court of Appeals are only binding for federal district courts within three states (Connecticut, New York, Vermont) and not for federal district courts located in other states. An analogous system of legal sources can be found at the state level in most U.S. states. The main principle of interaction between federal and state laws is that the rights that federal laws offer to individuals is a lower bound for state laws, i.e., states can grant their people more, but not fewer rights. Criminal law is a legal area reserved mostly for state laws, except for matters that cross the state borders, e.g., internet and airplane crimes, which fall under the federal jurisdiction \cite{samaha2016criminal}. Therefore, in the context of RRA, there exist two parallel sets of state and federal laws.
% \section{Synthesizing Theories: Analyze U.S. Sources of Law on RRA and Fairness}

% \label{sec:synthesize_theory}

% \subsection{Theoretical Findings}
% \label{subsec:theory_findings}
\section{Procedural Fairness and Outcome (Group and Individual) Fairness Are All Constitutionally Relevant}

% \section{First Challenge -- Relevancy of AI Fairness Criteria in Legal Sources}

% \tin{add more paragraph labels (Tracy's comment: Section 3.1 is really hard to read)}

% \subsection{Limited Adaptation of Individual Fairness from High to Low-level Legal Sources}

At a high level, the U.S. Constitution promotes fairness in criminal justice and other legal areas via two concepts: ``Due Process'' and ``Equal Protection''. Relevant excerpts are ``No person shall be [...] deprived of life, liberty, or property, without Due Process of law'' (the Fifth Amendment) and ``...nor shall any State deprive any person of life, liberty, or property, without Due Process of law; nor deny to any person within its jurisdiction the Equal Protection of the laws'' (the Fourteenth Amendment). Due process can be mapped to procedural (or process) fairness, e.g., which features should be used in an AI prediction model. Due to its straightforwardness, procedural fairness has been recognized by several low-level legal sources like statutes and regulations. As an example statute from the legislative branch, Section 3142(g) in the Bail Reform Act of 1984 specifies a long list of ``factors to be considered'' by judicial officers when making bail decisions, e.g., ``the history and characteristics of the person'' like employment and criminal history \cite{Bail_Reform_Act}. From the executive branch, several state regulations explicitly enumerate allowed ``risk factors'', e.g., age, sex/gender, number of prior convictions 
\cite{Pennsylvania_admin, California_regulations}, or arrests/charges/convictions that mandate sex-offender registration \cite{Oregon_admin}.

From the technical side, \citet{grgic2018beyond} seminally define procedural (process) fairness as “the fraction of all users who consider the use of every feature in $F$ [the set of input features] to be fair” and mathematically developed three procedural fairness metrics based on the aforementioned definition and further conditioning: “feature-apriori fairness” (users having no prior knowledge of how the feature usage affects outcomes), “feature-accuracy fairness” (users knowing that using the feature increases accuracy), and “feature-disparity fairness” (users knowing that using the feature increases outcome disparity). Our conceptualization of procedural fairness (whether a feature should be included in the input space of RRA tools) aligns with “feature-apriori fairness” by \citet{grgic2018beyond}.

However, procedural fairness compliance is difficult to evaluate, especially in the context of recidivism risk assessment tools as many models are proprietary, 
some of which are proprietary and only subject to experts' investigation of the algorithm if ordered by a judge, within the timeframe of a trial, and bound by several security protocols set by the tool owners, e.g., strict confidentiality constraints imposed by COMPAS owners on the expert witness Dr. Rudin in Flores v. Stanford (2021) \cite{2021flores}. In the recent RRA landscape, the majority of tools used by state governments come from private companies.\footnote{Legal Tech News (2020). The Most Widely Used Risk Assessment Tool in Each U.S. State. \url{https://www.law.com/legaltechnews/2020/07/13/the-most-widely-used-risk-assessment-tool-in-each-u-s-state/}.} 
Therefore, in the long run, courts may help facilitate procedural fairness standards in the RRA context by clarifying to what extent RRA tool deployers may invoke trade secret standards in criminal proceedings. For example, \citet{wexler2018life} argues that applying trade secret-based evidentiary privilege (to not disclose the technical inner-workings of RRA tools to the other parties during discovery), which were developed for the civil law context, directly to the criminal law context is “both harmful and unnecessary.” Furthermore, many theoretical rationales behind trade secret laws (e.g., to “facillitate controlled information sharing” with the government for public goods, and to incentivize owners of trade secrets to potentially turn them into disclosed patents) may in fact encourage trade secret disclosure to RRA decision subjects (defendants) through “narrow criminal discovery and subpoena powers combined with protective orders.” 

However, it seems more practical, at least in the short run, to evaluate outcome fairness, which might correspond to the ``Equal Protection'' clause in the Fourteenth Amendment because this clause is the cornerstone of many discussions against group-based classification and subordination \cite{balkin2003american}. 
As a reminder, within outcome fairness, the AI Fairness literature covers two main schools of fairness: group fairness and individual fairness.

In U.S. law, case law interprets the ``Equal Protection'' clause as covering both group fairness and individual fairness. Regarding group fairness, the U.S. Supreme Court rules in Washington v. Davis (1976) that group fairness (``disproportionate impact'') matters but there should be another relevant school of fairness: ``We have not held that a law [...] is invalid under the Equal Protection Clause simply because it may affect a greater proportion of one race than of another. Disproportionate impact is not irrelevant, but it is not the sole touchstone of an invidious racial discrimination forbidden by the Constitution'' \cite{1976washington}. As group fairness is `not the sole' criterion here, we trace back to older U.S. Supreme Court decisions to find the missing piece. For instance, F.S. Royster Guano Co. v. Commonwealth of Virginia (1920) interprets ``Equal Protection'' consistently with individual fairness if we define ``similarly situated'' as having a high similarity function score: ``The Equal Protection Clause of the Fourteenth Amendment commands [...] essentially a direction that all persons similarly situated should be treated alike'' \cite{1920fs}.

Royster is in the civil law context, so we examine U.S. Supreme Court's criminal law cases that also use the ``similarly situated [individuals]'' language when interpreting the constitutional ``Equal Protection'' concept: ``Under the Equal Protection component of the Fifth Amendment's Due Process Clause, [...] to establish a discriminatory effect in a race case, the claimant must show that \textit{similarly situated} individuals of a different race were not prosecuted'' (U.S. v. Armstrong, 1996 \cite{1996us}); ``The imposition of the death sentence upon petitioner pursuant to the new statute did not deny him Equal Protection of the laws. Having been neither tried nor sentenced prior to Furman, he was not \textit{similarly situated} to those whose death sentences were commuted'' (Dobbert v. Florida, 1977 \cite{1977dobbert}). This line of individual fairness interpretation for the `Equal Protection' clause is reaffirmed by more U.S. Supreme Court decisions such as Plyler v. Doe (1982). 
 
Interestingly, a synergy of the relevance of the ``Equal Protection'' clause to both group fairness and individual fairness is provided in City of Cleburne, Tex. v. Cleburne Living Center (1985) where the U.S. Supreme Court rules that ``Discrimination, in the Fourteenth Amendment sense, connotes a substantive constitutional judgment that two individuals or groups are entitled to be treated equally with respect to something'' \cite{1985cleburne}. Therefore, the ``Equal Protection'' clause in the U.S. Constitution has been interpreted as promoting both outcome fairness notions (group and individual fairness).

\section{Specific Policies for RRA Touch On Procedural and Group Fairness, but not Individual Fairness}

Athough some states such as California have developed elaborate legislation to control the quality of state-wide AI prediction tools with respect to group fairness metrics, we find limited evidence for individual fairness enforcement. Take for example Section 1320.35 of the California Penal Code, which from our search of AI-based pretrial risk assessment sources of law is the most fairness-detailed piece of legislation among U.S. states. However, most low-level fairness-related requirements in this statute are about group fairness only, requiring information about ``risk levels aggregated by race or ethnicity, gender, offense type, ZIP Code of residency, and release or detention decision'', ``the predictive accuracy of the tool by gender, race or ethnicity, and offense type'', and ``any disparate effect in the tools based on income level''\cite{California2021}. The evaluation of metrics such as risk levels (output rate), predictive accuracy, and impact across sensitive features (race or ethnicity, sex/gender) and proxy features (those that highly correlate with sensitive features) shows that California cares about mitigating not only direct bias but also indirect bias. For example, even if an AI model does not use the sensitive feature race, the real-world bias can still penetrate into the model's prediction via the use of features that highly correlate with race, e.g. ZIP Code of residence \cite{corbett2023measure}. 
Interestingly, an AI explainability angle is also present here as the statute requires validation information about ``line items, scoring, and weighting, as well as details on how each line item is scored'', or how much weight a prediction attributes to each feature. This explainability requirement exposes the inner decision-making process of the model, so it can be linked to procedural fairness. However, individual fairness is not relevant in this statute. 
% Both group (outcome) fairness and, to a lesser extent, procedural fairness criteria can be identified in the statute. However, the closest requirement we find that is remotely related to individual (outcome) fairness is the option to share individual-level data conditioned on research contracts. This option might be either to evaluate an individual fairness metric or to perform many other types of evaluation including the computation of other group fairness metrics, e.g., false positive or false negative. Therefore, individual fairness is not directly relevant to this statute. 
Another example is Section 725 ILCS 5/110-6.4 of the Illinois Compiled Statutes. It requires their state-wide risk assessment tool to ``not discriminate on the basis of race, gender, educational level, socio-economic status, or neighborhood'', which can only be attributed to group fairness, not individual fairness \cite{Illinois2023}. 

Procedural fairness considerations can also be found in regulations (executive branch). For example, 204 Pa. Code § 305.2 \cite{Pennsylvania2019} explicitly supports using age and sex/gender as RRA model inputs. 

\begin{figure}
\centering
    \includegraphics[width=0.5\textwidth]{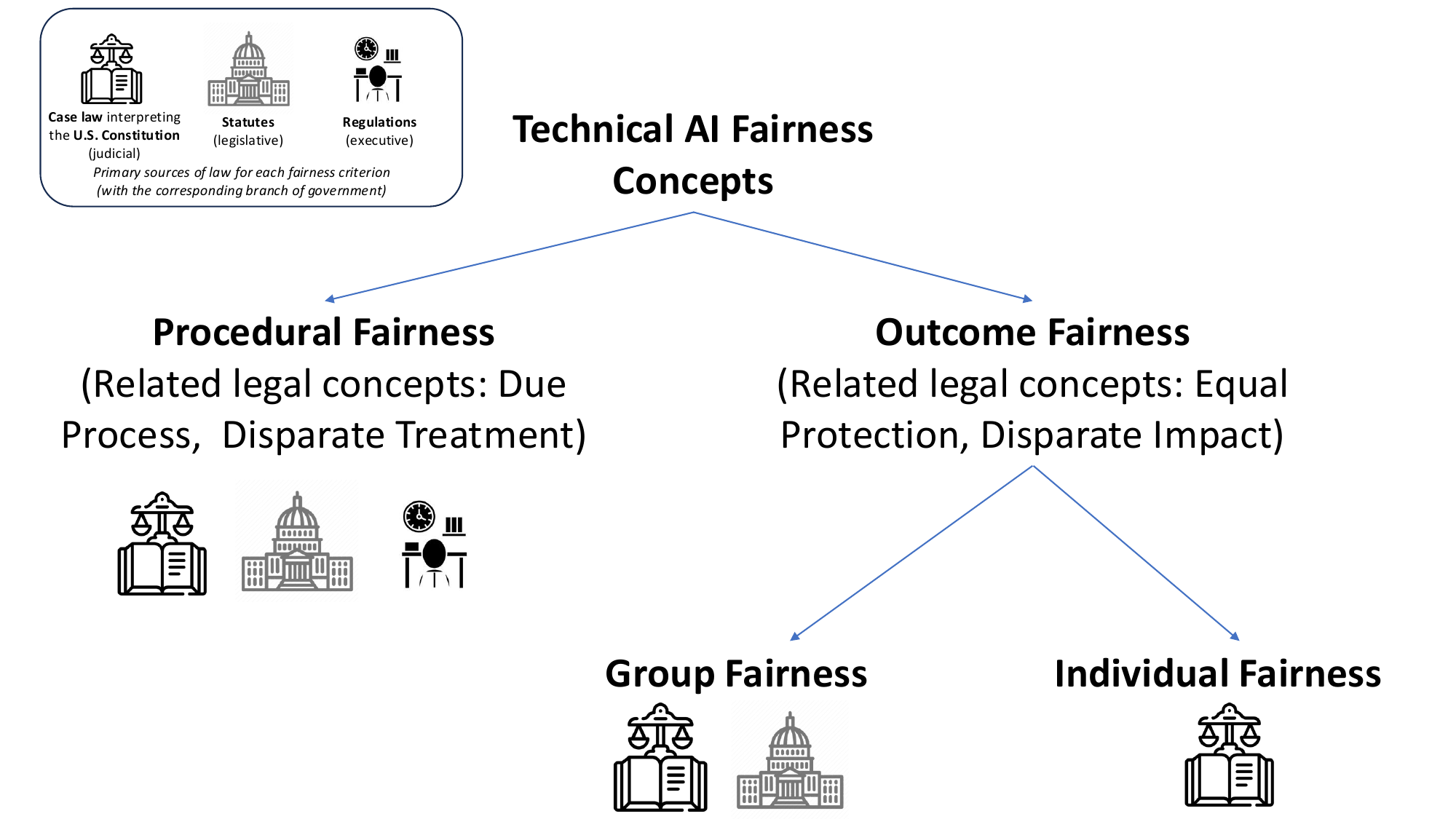}
    \caption{Whether Individual, Group, and Procedural Fairness Concepts are recognized by the U.S. Constitution, case law (judicial), statutes (legislative), or regulations (executive)}
    \label{group_vs_individual}
\end{figure}

In summary, while the three technical fairness concepts (procedural, group, and individual fairness) may have a constitutional basis, among them, only procedural fairness and group fairness can be found in specific statutes and regulations, as shown in Figure
\ref{group_vs_individual}.
%Figure \ref{group-vs-individual}.

%Note: the star (*) makes sure the figure does not overlap with text

\section{Demographic Features in the Same Judicial ``Scrutiny'' Category Might Trigger Slightly Different Scrutiny by U.S. Courts in Practice}
%\tin{add more examples to better show disagreement?}

% An important first implementation step in enforcing either individual or group fairness criteria is to determine which features are appropriate to use as part of the audit pipeline, either to compute the similarity function between any two individuals (for individual fairness) or to evaluate a metric that matters across different groups with respect to each of those features (for group fairness). 

As procedural fairness is found in concrete statutes and regulations, it is not surprising that the relationship between demographic features and procedural fairness has been well established in case law. For example, in Clark v. Jeter (1988), the U.S. Supreme Court acknowledges a common law hierarchy of features (classes or group memberships) on which the government may discriminate to varying degrees for the sake of public interest, from the most to the least stringent amount of justification the government must provide: ``strict scrutiny'' (e.g., race and alienage/national origin), ``intermediate scrutiny'' or ``heightened scrutiny'' (e.g., sex/gender and legitimacy/out-of-wedlock status), and ``rational basis'' (e.g., age and disability) \cite{1988clark}. 

In particular, at one end of the judicial scrutiny-worthiness spectrum, a ``strict scrutiny'' feature means that a state or federal government's policy that treats one group differently than another group based on that feature (e.g., race) must be the “least restrictive means” to further a ``compelling government interest."\footnote{\url{https://www.law.cornell.edu/wex/strict_scrutiny}}
At the other end of the spectrum, a ''rational basis'' feature means that a government's differential treatment policy based on that feature (e.g., age) must serve a ``legitimate state interest'', and there must be a ``rational'' connection between the policy's means and interest.\footnote{\url{https://www.law.cornell.edu/wex/rational_basis_test}}
In the middle of the spectrum, an ``intermediate/heightened scrutiny'' feature means that government's differential treatment policy based on that feature (e.g., sex/gender) must further an ``important government interest'' (a higher burden of proof than the ``legitimate'' interest in rational basis, but lower than the ``compelling'' interest in strict scrutiny), and the policy's means must be ``substantially related to'' that interest (a higher burden of proof than a ``rational connection'', but lower than the ``least restrictive means'').\footnote{\url{https://www.law.cornell.edu/wex/intermediate_scrutiny}}

Instead of viewing strict scrutiny, intermediate scrutiny, and rational basis as three discrete values of warranted scrutiny amount, we may instead formulate them as three continuous ranges because case law has shown that two features within the same group (or range) may not necessarily warrant the exact same amount of scrutiny. 

For example, U.S. Supreme Court’s (SCOTUS) decisions recognize race and alienage as strict scrutiny features: 1/ Korematsu v. United States, 323 U.S. 214 (1944): ``all legal restrictions which curtail the civil rights of a single racial group are immediately suspect. [...] courts must subject them to the most rigid scrutiny.'' \cite{1944korematsu}. 2/ Graham v. Richardson, 403 U.S. 365 (1971): ``classifications based on alienage, like those based on nationality or race, are inherently suspect and subject to close judicial scrutiny'' \cite{1971graham}. 3/ Bernal v. Fainter, 467 U.S. 216 (1984): ``a state law that discriminates on the basis of alienage can be sustained only if it can withstand strict judicial scrutiny'' \cite{1984bernal}.

However, case law shows that SCOTUS has scrutinized alienage-based discrimination with two slight de facto differences compared to race-based discrimination.

The first difference is that while SCOTUS mandates the same amount of (strict) scrutiny applies regardless of whether a race-based discrimination comes from a state or federal government, SCOTUS applies less scrutiny on alienage-based discrimination from the federal government than on alienage-based discrimination from state governments. In Hampton v. Mow Sun Wong (1976), SCOTUS hints that judicial scrutiny over alienage-based discrimination by the federal government in civil service hiring may be lower than scrutiny over alienage-based discrimination by a state government: “overriding national interests may justify a citizenship requirement in the federal service even though an identical requirement may not be enforced by a State” \cite{1976hampton}. 
However, for race-based discrimination, in Adarand Constructors, Inc. v. Peña, 515 U.S. 200 (1995), SCOTUS reasons that “All racial classifications, imposed by whatever federal, state, or local governmental actor, must be analyzed by a reviewing court under strict scrutiny’’ \cite{1995adarand}. 

The second difference is a “political function” exception, which enables state/federal government to justify alienage-based discrimination but not race-based discrimination. In Bernal v. Fainter, 467 U.S. 216 (1984), “We [SCOTUS] have, however, developed a narrow exception to the rule that discrimination based on alienage triggers strict scrutiny […] and applies to laws that exclude aliens from positions intimately related to the process of democratic self-government’’ \cite{1984bernal}. 
SCOTUS used this “political function” exception to avoid striking down state policies that exclude aliens (non-U.S. citizens) from public service jobs such as 
police officers in Foley v. Connelie \cite{1978foley}, 
teachers in Ambach v. Norwick \cite{1979ambach}, 
and probation officers in Cabell v. Chavez-Salido \cite{1982cabell}.  

In conclusion, case law supports our decision to model each scrutiny standard (e.g., strict scrutiny) as a continuous-valued range (instead of a discrete category) to accommodate several demographic features that may slightly differ in de facto scrutiny-worthiness (e.g., alienage and race). 

\section{Jurisdictions Disagree on Procedural Fairness Criteria (Which Demographics to Use as Model Inputs)}

Although the scrutiny-based hierarchy is not legally binding for the recidivism context, there is consensus not to use features in the ``strict scrutiny'' range as model inputs. The consensus on race is clear. Although most risk assessment tools, e.g., COMPAS, do not use race as a feature, many works such as \citet{johnson2021algorithmic} go further to criticize the tool for indirectly perpetuating racial bias through proxy features (those highly correlated with race). \citet{hu2023what} gives a constructivist argument for why acting on proxy features of race is not different than acting on the basis of race.
Regarding alienage/national origin, a sample questionnaire\footnote{\url{https://www.documentcloud.org/documents/2702103-Sample-Risk-Assessment-COMPAS-CORE}} shows that COMPAS does not collect this feature, and we do not find any literature discussing the use of or unfairness with respect to this feature by any recidivism tools, indicating an implicit consensus that national origin should not be used.

For the lower two ranges (intermediate scrutiny and rational basis), there remains disagreement on whether features at those ranges might be used. The first example of disagreement is sex/gender (at the intermediate scrutiny range). On the one hand, the Wisconsin Supreme Court in their State v. Loomis (2016) decision strongly advocates the use of sex/gender as a feature: ``COMPAS's use of gender promotes accuracy that ultimately inures to the benefit of the justice system, including defendants'' \cite{2016state}. 
On the other hand, Section 2A:162-25(2) of the New Jersey Statutes adopts a clear stance against using sex/gender as a feature for AI recidivism risk assessment: ``Recommendations for pretrial release shall not be discriminatory based on race, ethnicity, gender, or socio-economic status'' \cite{NewJersey}. 

The second example of disagreement is age (at the rational basis range). For example, in the same jurisdiction of New York, while Section 168-l of the Consolidated Laws of New York explicitly requires sex offense recidivism risk assessment to take into account ``the age of the sex offender at the time of the commission of the first sex offense'' \cite{NewYork2011}, in their Flores v. Stanford (2021) decision, the U.S. district court for Southern District of New York indicates an implicit stance against the use of age in recidivism risk assessment by allowing expert inspection of the data used to train COMPAS to ``help Plaintiffs substantiate their allegations that COMPAS punishes juvenile offenders for their youth, such that Defendants' reliance on this tool is constitutionally problematic'' %as the Court believes that `the weight COMPAS affords to youth in determining recidivism risk will be highly relevant to the underlying constitutional claims' 
\cite{2021flores}.

Not only case law and statutes, but state regulations also disagree. While some regulations include both sex/gender and age as risk factors (e.g., 204 Pa. Stat. Ann. § 305.2 \cite{Pennsylvania2019} and Cal. Code Regs. tit. 15, § 3768.1 \cite{California_regulations}), other regulations include neither sex/gender nor age (e.g., 20 Ill. Adm. Code 1905.60 \cite{Illinois_admin} and Or. Admin. R. 291-078-0020 \cite{Oregon_admin_probation}). 

In summary, while there is a consensus that strict scrutiny features should be excluded from model inputs, legal sources disagree on intermediate scrutiny (sex/gender) and rational basis (age) features.
\section{Proposal: Integrating Fairness Criteria and Demographic Features into a Scrutiny-based Framework}
%\subsection{How much Scrutiny and Which Fairness Criteria does each Feature Warrant?}
\label{sec:extend_theory}

The controversy on whether features in the two lower ranges of scrutiny can be used as input for recidivism risk assessment models indicates room for further research that links to the procedural vs. outcome fairness and the individual versus group fairness dichotomies. In the same framework of three continuous scrutiny ranges, we propose three discrete scrutiny thresholds: First is the procedural fairness or ``exclusion from model inputs'' threshold (if we take the view by \citet{lee2019procedural, agan2018ban} that excluding a feature from the model input space corresponds to procedural fairness). Second is the group fairness or ``group parity required'' threshold. Third is the individual fairness or ``ignorance in individual similarity function'' threshold. The more descriptive (and lengthier) name of each threshold corresponds to a pass condition, i.e., a feature's scrutiny amount warranted is higher than this threshold if people believe that the feature should satisfy the corresponding threshold's pass condition.

First, a feature that people believe should be excluded from the input space of an RRA model will have its scrutiny score passing (above) the procedural fairness threshold. Second, for a feature, if people believe it should be the reference feature so that they can compute a group parity metric and require that parity metric to be near perfect, then this feature's scrutiny score will pass the group fairness threshold. Finally, if people believe that to evaluate an individual fairness metric, the individual similarity function must ignore a certain feature, then this feature's scrutiny score will pass the individual fairness threshold.

\begin{figure}
\centering
    \includegraphics[width=0.5\textwidth]{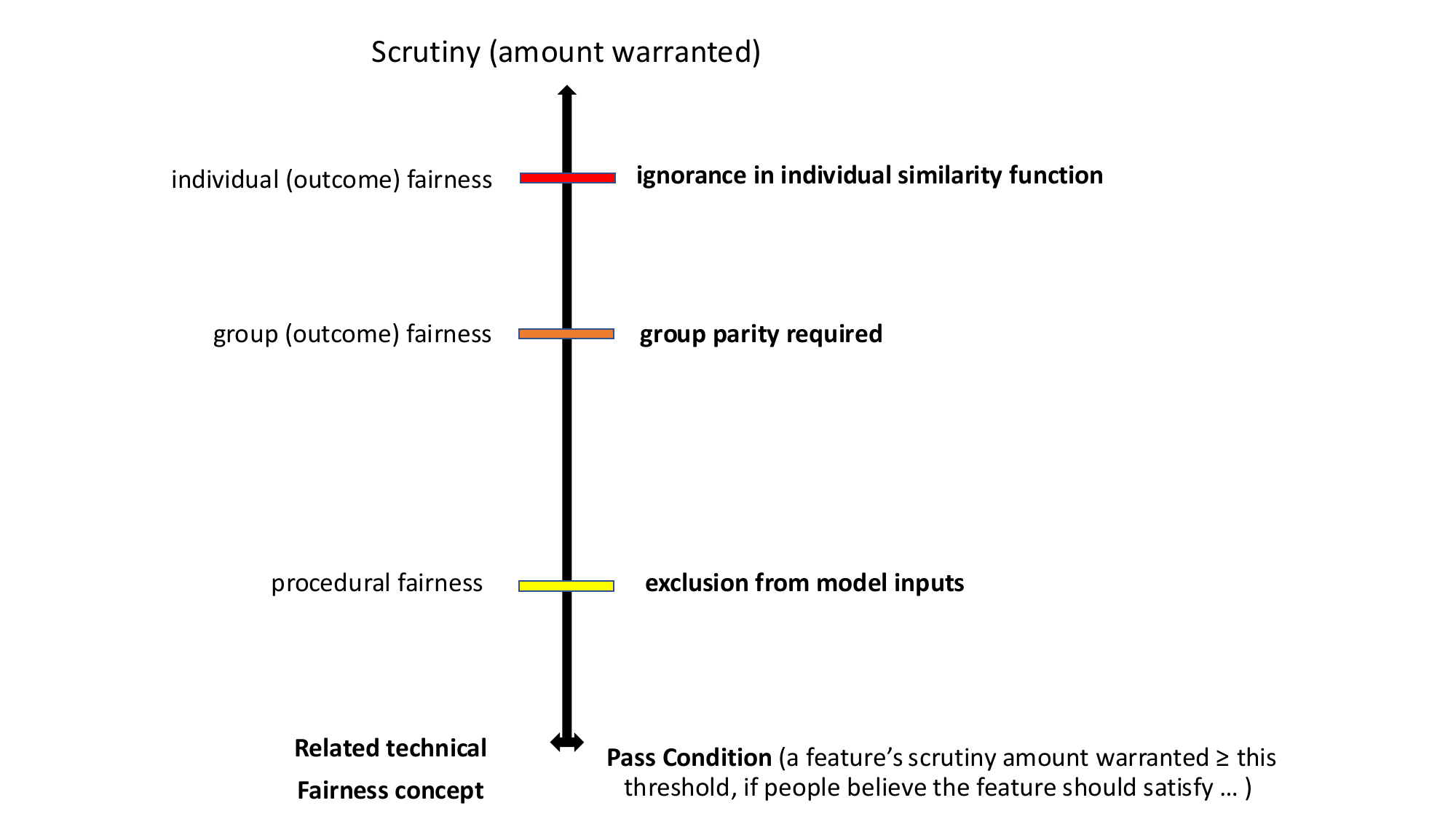}
    \caption{Potential ordering of proposed scrutiny thresholds.}
    \label{fig-ordering}
\end{figure}

It remains for future work to investigate whether it is normatively plausible to, and if so, how to rank these three fairness thresholds against one another. A potential, unvalidated ordering of them is illustrated in Figure \ref{fig-ordering}. Note that satisfying a higher threshold does not mathematically guarantee satisfying a lower threshold, e.g., perfect outcome parity across races does not guarantee that race is excluded from model inputs. Rather, any potential ordering is only a normative judgment on which threshold should reflect higher judicial scrutiny. 

We provide some justifications for this not yet empirically validated ordering: 
An example of why the group fairness threshold (orange) is above the procedural fairness threshold (yellow) is the recurrent debate on whether COMPAS should achieve racial parity (i.e., race may pass or fail the orange threshold) even though COMPAS does not use race as a predictive feature (i.e., race passes the yellow threshold) \cite{washington2018argue}. 
An intuition for why the individual fairness threshold (red) is above the group fairness threshold (orange) is a thought experiment: even if one believes that a model should not have disparate racial impact (i.e., race passes the orange threshold), one might still believe that all else equal, two same-race individuals are more similar than two different-races individuals (i.e., race fails the red threshold). Thinking reversely, suppose one believes that race is not a good proxy to measure criminal/violence prevalence (or equivalently, all races have the same amount of criminal/violence prevalence), and therefore, race should be excluded from the pairwise individual similarity function (i.e., race passes the red). This premise should imply that the recidivism risk should be independent of race or equalized across racial groups (i.e., race passes the orange threshold). In other words, passing the red threshold might imply normatively, but not mathematically, passing the orange threshold, indicating that the individual fairness threshold reflects higher scrutiny than the group fairness threshold.
However, this ordering has not been empirically validated.

%An illustration of our three proposed scrutiny thresholds is given in Figure \ref{fig1}. 

\begin{figure}
\centering
    \includegraphics[width=0.5\textwidth]{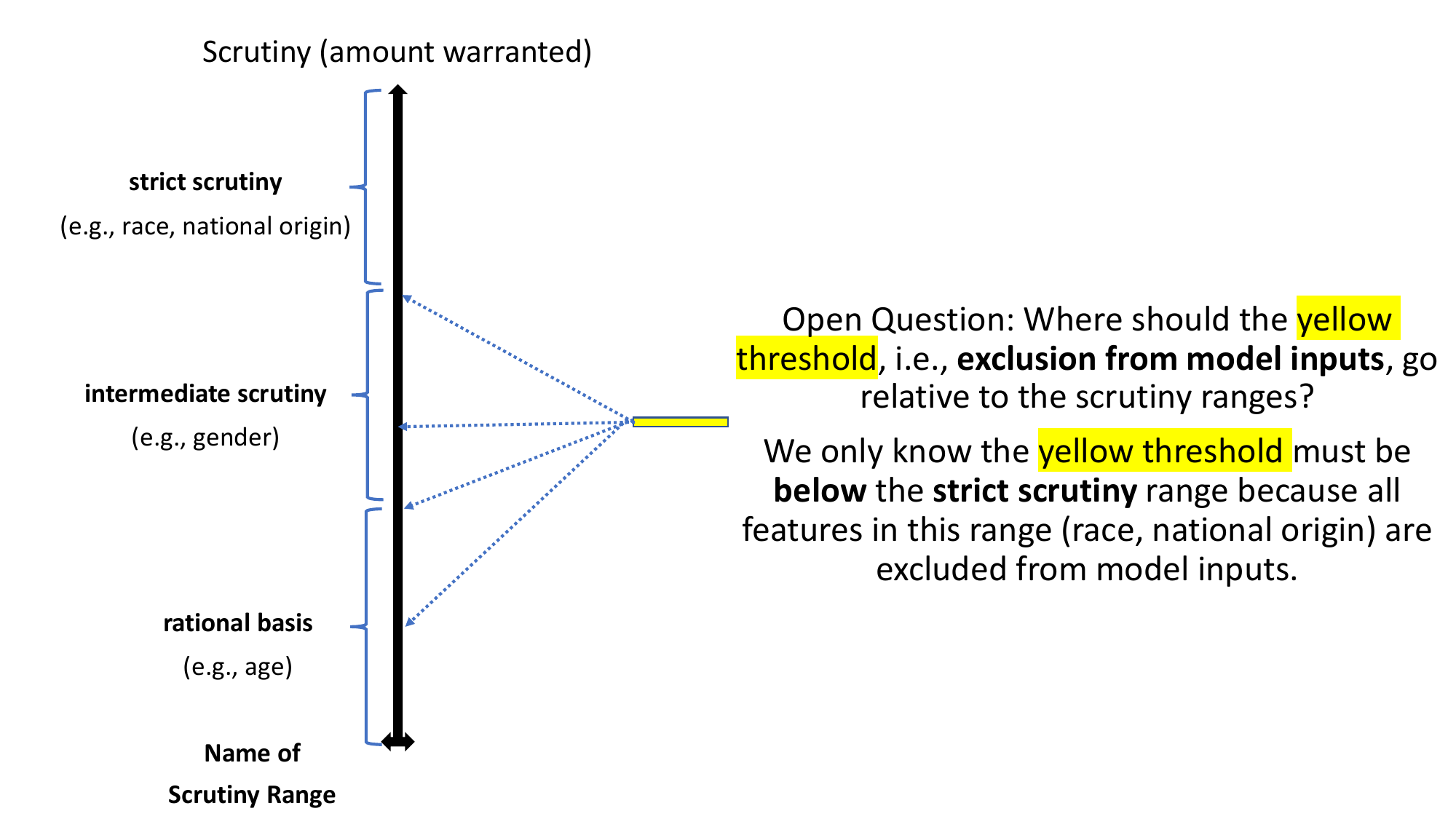}
    \caption{Legal sources classify demographic features into three scrutiny ranges. For RRA, only strict scrutiny features are above the ``exclusion from model inputs'' (yellow) scrutiny threshold, i.e., expected by legal sources to be excluded from the AI models' input space. There are conflicting legal sources about whether intermediate scrutiny or rational basis features (such as sex/gender or age) should be excluded.}
    \label{scrutiny}
\end{figure}

Contextualizing our fairness thresholds in the judicial scrutiny-based framework, we only know that the procedural fairness threshold must be below the strict scrutiny range because all features in this range (race, national origin) are excluded from model inputs. As legal sources disagree on whether sex/gender and age should be excluded from model inputs, we do not know where our procedural fairness threshold should go with respect to the intermediate scrutiny and rational basis ranges. We illustrate this open question in Figure \ref{scrutiny}. %in \autoref{appendix-results}. %(Appendix A). %Once we know where our procedural fairness threshold should be relative to the three scrutiny ranges, follow-up human study and/or legal research can be conducted to determine where our group fairness and the individual fairness thresholds should be on the scrutiny spectrum.

\begin{figure}
\centering
    \includegraphics[width=0.5\textwidth]{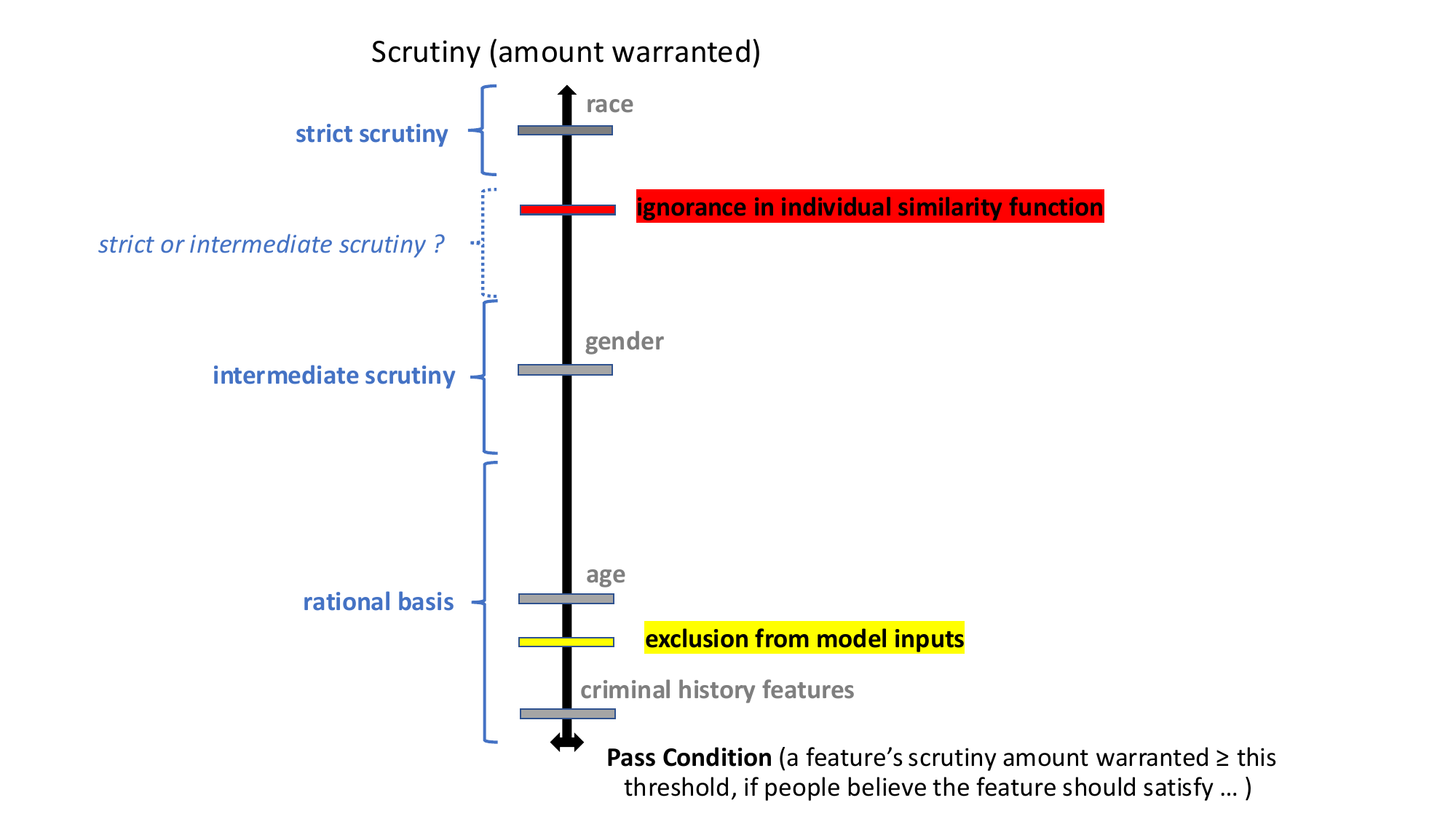}
    \caption{How the empirical findings by \citet{nguyen2025demographic} might be applied to our scrutiny-based framework.
    }
    \label{fig:scrutiny-results}
\end{figure}

Suppose empirical works, e.g., human studies with laypeople or court officials, can confirm that our proposed three scrutiny thresholds are consistent with humans' fairness perception, our framework might help systematically determine whether procedural, group, and/or individual fairness criteria should be audited with respect to each demographic feature. 
% For example, if both crowdworkers and experts come to a consensus that age should be used as an input feature for recidivism risk assessment (i.e. age is below or fails the procedural fairness threshold and thus the group fairness and the individual fairness thresholds of scrutiny too), our framework indicates that parity with respect to age groups does not matter, but age can be used to compute the similarity function for individual fairness.

One empirical work highly relevant to our framework is the human subjects experiment conducted by \citet{nguyen2025demographic}, which tests whether including a demographic feature (race, sex/gender, or age) as available information for laypeople to evaluate pairwise similarity between criminal justice defendants would worsen those laypeople's individual fairness judgement (i.e., making them more likely to judge opposite risk scores on similar defendants as fair). They conclude that the pairwise individual similarity function should include sex and age but ignore race. They also have a short survey question on procedural fairness and find that most laypeople prefer demographic features (race, sex/gender, and age) to be excluded from the RRA model inputs.

Figure \ref{fig:scrutiny-results} summarizes how the human subjects experiment by \citet{nguyen2025demographic} might further clarify our proposed framework. 
Regarding the ``exclusion from model inputs'' (yellow) threshold, most of their participants believe criminal history features should be included (i.e., below the yellow threshold) and a majority of participants think that age, sex/gender, and race should be excluded (i.e., above the yellow threshold). 
Regarding the ``ignorance in individual similarity function'' (red) threshold, their empirical results and analyses conclude that the pairwise individual similarity function can consider age and sex/gender (i.e., below the red threshold), but it should ignore race (i.e., above the red threshold). Note that due to the relative positions of the red and yellow thresholds compared to sex/gender and age (red $>$ \{ sex/gender, age \} $>$ yellow), applying the transitive property of inequality, the red threshold is above the yellow threshold, partially validating our theoretically proposed ordering. Finally, the three (blue) scrutiny ranges from legal precedents (strict scrutiny, intermediate scrutiny, rational basis) are placed around each respective demographic feature. However, their results do not confirm whether the red threshold falls within the strict or intermediate scrutiny range.

As a caution, this is just one out of many possible empirical arrangements of our proposed framework. For example, another possible arrangement is that sex/gender and age be placed below the yellow (procedural fairness) threshold because \citet{katsiyannis2018adult} found sex/gender and age to be predictive of recidivism. Therefore, we invite future empirical works to refine our judicial scrutiny-based framework of fairness criteria and demographic features.

\section{International Perspectives: How our U.S. Law Findings Relate to Other Major Jurisdictions}

We contextualize our U.S. legal research findings and proposed scrutiny-based framework in other jurisdictions where there is a strong AI market and evidenced/potential use of AI-assisted decision-making tools by government bodies. 

\subsection{The European Union}
The EU AI Act categorizes RRA tools ("Individual criminal offence risk assessment or prediction") as ``unacceptable risk'' and therefore bans it. However, according to Recital 42: ``risk assessments carried out with regard to natural persons in order to assess the likelihood of their offending or to predict the occurrence of an actual or potential criminal offence based solely on profiling them or on assessing their personality traits and characteristics should be prohibited. In any case, that prohibition does not refer to or touch upon risk analytics that are not based on the profiling of individuals or on the personality traits and characteristics of individuals, such as AI systems using risk analytics to assess the likelihood of financial fraud by undertakings on the basis of suspicious transactions'' \cite{eu_ai_act_2024}. This likely means that any AI-assisted decision making tools used by EU member states are prohibited from including any demographic features (e.g., race, sex, or age) as model input features, i.e., all demographic features are placed above the ``exclusion from model inputs'' threshold if we adapt our scrutiny-based framework to the EU context.

Prior to the EU AI Act, as reviewed by \citet{weerts2023algorithmic}, four main directives lay out the theoretical framework for EU non-discrimination law, which also applies to algorithmic fairness considerations: 
the Race Equality Directive 2000/43/EC \cite{racial_equality_directive}, 
the Framework Equality Directive \cite{framework_equality_directive},
the Gender Equality Directives 2004/113/EC \cite{gender_equality_directive_2004}
and 2006/54/EC \cite{gender_equality_directive_2006}. 
This EU framework defines two types of fairness concerns: direct discrimination ("one person is treated less favourably than another is, has been or would be treated in a comparable situation on grounds of" a protected feature like race, sex/gender, and age) and indirect discrimination (``an apparently neutral provision, criterion or practice would put persons of a protected group at a particular disadvantage'').
While the strictly forbidden ``direct discrimination'' directly reaffirms our procedural fairness observation from the EU AI Act that race, sex, or age should be excluded from model input features, the less restricted ``indirect discrimination'' under EU non-discrimination law hints at our ``group parity required'' scrutiny threshold. Interestingly, ``indirect discrimination'' might be legal if it ``is objectively justified by a legitimate aim and the means of achieving that aim are appropriate and necessary", which was interpreted as ``if the same legitimate aim can be achieved through less discriminatory alternatives, those must be used'' \cite{weerts2023algorithmic, tobler2005indirect}. 
This ``[no] less discriminatory alternatives'' language of permissible ``indirect discrimnation'' under EU non-discrimination law is highly similar to the ``least restrictive means'' language in the U.S. ``strict scrutiny'' review standard against discriminatory government policies. If an AI model with disparate impact (e.g., violating demographic parity against certain races) can survive ``indirect discrimination'' challenges under EU law, it is likely to survive the ``strict scrutiny'' test under U.S. law too.

Regarding individual fairness, \citet{weerts2023algorithmic} find the ``likes should be treated alike'' Aristotelian formulation of justice to be the philosophical foundation of the EU ``direct discrimination'' doctrine. However, after thoroughly reviewing specific and operationalizable AI-related sources of EU law such as 
the General Data Protection Regulation, GDPR \cite{gdpr_2016}, 
the Digital Services Act, DSA \cite{dsa_2022},
and a proposal version of the EU AI Act, 
\citet{calvi2023enhancing} conclude that ``the few references to fairness metrics existing in the EU legal framework hint at group fairness and not individual fairness.'' This problem is similar to our finding that 
although SCOTUS linked the U.S. constitutional ``Equal Protection'' clause to the technical individual fairness concept, we find no concrete individual fairness standards in U.S. federal/state statutes or regulations.

\subsection{China}
Although Chinese courts do not have an RRA use case, some of their criminal courts use AI to estimate the ``social harm'' of the current case based on similar past cases from other defendants (instead of predicting future criminal cases from the same defendant). Judges may consider AI-estimated ``social harm'' as ``a frame of reference for sentencing,'' which is analogous to U.S. judges considering RRA risk scores in sentencing \cite{papagianneas2023fairness, cui2020artificial}.

Regarding technical/legal fairness criteria, unlike the procedural vs. outcome fairness dichotomy in American/European laws, \citet{papagianneas2023fairness} synthesize a procedural vs. substantive fairness dichotomy after reviewing judicial documents published by China's Supreme People's Court (SPC) to guide lower courts. Substantive fairness in China has nothing to do with outcome-based (group/procedural) fairness, but its goals are 1/ sustaining legitimacy (of the ruling party), 2/ maintaining social stability, and 3/ user convenience. On the other hand, procedural fairness in China is often less about a set of legally allowed/forbidden features for model inputs, but more about the explainability angle (internal accountability, procedural consistency, and external visibility), e.g., whether AI-assisted decision makers may explain to decision subjects what features they consider and why. Furthermore, the expected response to potentially discriminatory actions by a government agency in China is often to report to that agency or a higher government body for a top-down solution, rather than through bottom-up litigation against the government, i.e., judicial scrutiny of government actions is rare \cite{cui2019judicial}.
Therefore, our judicial scrutiny-based fairness framework is irrelevant to China's legal system.

\subsection{India}
The most important supplement to our framework in India's context is to expand the set of demographic features for fairness consideration to include ``caste'', a hereditary Hindu class feature which historically determines one's access to financial, cultural, social, and many other types of capital in life \cite{sambasivan2021re}. 
If placed on our legal framework, caste and sex/gender are likely above the ``group parity required'' threshold due to ``reservations'', a constitutionally mandated practice to reserve certain quotas to access public resources like education and bank loans for vulnerable groups such as historically marginalized castes and women \cite{basavaraju2009reservation, sambasivan2021re}. This policy seeks to equalize access to resources across demographic groups and is therefore related to group fairness. 
A recent case decided by the Supreme Court of India, LT. Col Nitisha v. Union of India (2021), though not in the AI context, further reaffirms group fairness protection for women as the Court recognizes indirect discrimination by the government (e.g., disparate impact of a policy on women, even without proven discriminatory intent) to be unconstitutional, just like direct discrimination \cite{nitisha2021}.

% Caste, sex/gender, and also race are likely be above the ``exclusion from model inputs'' threshold thanks to Article 15(1) of India's Constitution: ``The State shall not discriminate against any citizen on grounds only of religion, race, caste, sex, place of birth or any of them''.
\section{Discussion}

We contemplate how our legal findings and new framework might impact different stakeholders in the design and refinement of RRA and other AI-assisted decision-making tools.

\paragraph{Tool designers}
Though not directly in the criminal justice context, some HCI scholars and practitioners started to consider fairness as part of the design pipeline. For example, \citet{nakao2022toward} designed an interactive ``explanatory debugging'' interface to help non-expert end-users identify and potentially correct fairness issues in the loan application context. However, this is still a minority. For example, within the Mobile HCI community, \citet{yfantidou2023state} expressed concerns that only 5\% of the Mobile HCI papers they reviewed conformed to fairness criteria, calling for more systematic fairness research in that sub-community. 
In HCI more broadly, some designers of AI-assisted decision-making tools like RRA might not have enough incentives to spend extra time and resources on auditing any potential bias in their products or mitigating discovered bias through algorithmic refinement or alternative data. 
Our legal research might provide designers with external, legally binding incentives to adopt fair and responsible AI practices, as \citet{grimpe2014towards} showed that policy might facilitate the adoption of responsible design principles (e.g., inclusiveness) into design practices.

Some other designers might simply not have enough knowledge about the myriad of technical fairness criteria or fair design principles out there in the technical AI literature. By integrating technical fairness criteria and demographic features into a framework of scrutiny, we might streamline the fairness audit process to help designers understand what types of demographic data they need to collect and what types of fairness criteria they should optimize for.

\paragraph{Tool users}
When a high-stakes AI tool makes bad (e.g., unfair) decisions, it is important to enable the tool users to recognize and correct such decisions. This problem of Algorithmic Recourse has gained increasing attention in HCI, especially Human-centered AI. For example, \citet{esfahani2024preference} designed an interaction paradigm to elicit users' preferences for algorithmic decisions and guide users towards effective recourse interventions for unfavorable decisions. In RRA and other high-stakes contexts, the users are often specialized experts who follow specific professional traditions, which algorithm recourse should also consider. For example, \citet{yacoby2022if} found that counterfactual explanations, a popular algorithmic recourse to uncover bias \cite{karimi2021algorithmic}, were simply ignored by judges because judges perceived their job as deciding real cases for real people, not hypothetical ones. 

%\tin{cite hal's upol paper}
Judges have little incentive to consider algorithmic recourse or avoid overreliance on RRA tools because, as long as a judge still makes the final bail/sentencing decision, it becomes a shield to protect their decision against accusations of fairness violation in future appeals. In Brooks v. Commonwealth, although there is evidence that the trial judge favors the AI risk assessment-based sentencing recommendation and dismisses the shorter active sentence in the non-AI recommendation, the Court of Appeals of Virginia rules that `the trial court properly exercised its discretion' without inquiring the level of reliance by the trial court on the AI risk assessment \cite{2004brooks}.

Therefore, by attributing a strong legal basis to existing technical fairness criteria and our proposed scrutiny-based framework, we encourage future work on designing new recourse intervention methods based on these criteria and framework. Such legally grounded recourses might confront the judges with evidence of potential bias in the RRA tool more directly, incentivizing them to respond to such bias more promptly. For example, if a recourse can show that an RRA tool gave opposite risk scores to many pairs of defendants with different races (but similar criminal histories, sexes/genders, and ages), which we showed to violate the constitutionally relevant individual fairness criterion, judges will be more likely to suspend the use of that tool in their court rooms and require further fairness audits.

\paragraph{Tool decision subjects}

Before one can find recourse for bad algorithmic decisions, one must be able to contest those. The contestability literature in HCI often assumes the one who should contest is the model user \cite{vaccaro2019contestability}. However, \citet{lyons2021conceptualising} found that first, the primary goal of contestability is to protect individuals, and second, contestability in AI-assisted decision-making resembles contestability in human decision-making. Therefore, in RRA and other high-stakes contexts where the decision subjects (e.g., defendants) differ from the model users (e.g., judges), the primary focus of contestability design should be on the decision subjects because they have the most incentives to protect themselves and are separate from the decision makers, i.e., tool users (so that research on AI-assisted decision-making contestability can draw analogies from human decision-making contestability).

RRA decision subjects face a high imbalance of contestability power against RRA users because the burden of proof to show that judges over-rely on AI is often placed on the defendants. In People v. Younglove (2019), when the defendants contended that their right to an `individualized sentencing decision' was violated by the COMPAS risk score presented to the sentencing judge, the Michigan Court of Appeals ruled that ``defendants offer no evidence that their sentencing courts actually placed significant (or any) weight on the COMPAS assessments in crafting their sentences. Defendants have failed to carry their burden of showing that the inclusion of the information affected their substantial rights'' \cite{2019people}.

Defendants have almost no way to reconstruct the mental model of the judges' decisions, especially when most judges are not required to disclose how much weight they give to the AI risk score, which may internalize demographics-based bias.
For example, \citet{stevenson2022algorithmic} showed that, conditioned on the same recidivism risk level, AI-assisted judges in Virginia give black defendants sentences that are 15-20\% longer than white defendants. 

By integrating nuanced legal concepts of scrutiny and abstract fairness criteria into a simple framework, we might empower criminal defense attorneys to spot fairness problems of RRA tools used on their clients more effectively, thereby enhancing the clients' contestability and potentially suing the tool designers for wrongful discrimination or personal liberty damages, which may incentivize fairer algorithmic design and refinement. An attorney is often the closest companion a defendant gets when going through the criminal justice system. \citet{karusala2024understanding} found that accompaniment, i.e., being accompanied by a professional who also cares about the decision subject, is critical for marginalized communities (those most susceptible to algorithmic bias) to contest AI decisions. 

\paragraph{How Technical AI Fairness Works Further Inform Our Legal Framework}
\citet{rudin2019stop} argues against using black-box RRA tools because even if one tries to interpret black-box models’ predictions with post hoc explanations (developing a second model to perturb the original model’s outcomes to explain it), those explanations are not necessarily faithful. Their advocating for inherently interpretable models in high-stakes contexts like RRA will enable easier audit of “procedural fairness” in our framework, as expert witnesses might investigate the source code to determine whether the model uses a demographic feature and with how much weight.
Both works by \citet{lagioia2023algorithmic} and \citet{chouldechova2017fair} demonstrate the potential incompatibility of several group fairness metrics (“a system that is equally accurate for different groups may fail to comply with group-parity standards”; “disparate impact can arise when an RPI fails to satisfy the criterion of error rate balance”), informing our future work to explore other variations of our group fairness-based (yellow) scrutiny threshold and their relative orders in our framework.

\paragraph{Conclusion}
Our legal research shows that while major technical AI fairness criteria (procedural, group, and individual fairness) have constitutional bases, they have not yet been translated into consistent RRA statutes and regulations. To fill this gap, we extend the demographics-related scrutiny framework from U.S. case law to incorporate fairness criteria as scrutiny thresholds. Finally, we contextualize our framework in three other jurisdictions with a big AI market.

%AIES_2025__law_review/
\bibliography{aaai25}

\begin{thebibliography}{104}
\providecommand{\natexlab}[1]{#1}

\bibitem[{Agan and Starr(2018)}]{agan2018ban}
Agan, A.; and Starr, S. 2018.
\newblock Ban the box, criminal records, and racial discrimination: A field experiment.
\newblock \emph{The Quarterly Journal of Economics}, 133(1): 191--235.

\bibitem[{Angwin et~al.(2022)Angwin, Larson, Mattu, and Kirchner}]{angwin2022machine}
Angwin, J.; Larson, J.; Mattu, S.; and Kirchner, L. 2022.
\newblock Machine bias.
\newblock In \emph{Ethics of data and analytics}, 254--264. Auerbach Publications.

\bibitem[{{Bail Reform Act}(1984)}]{Bail_Reform_Act}
{Bail Reform Act}. 1984.
\newblock Release or detention of a defendant pending trial.
\newblock ({18 U.S.C. {\S} 3142}).

\bibitem[{Balkin and Siegel(2003)}]{balkin2003american}
Balkin, J.~M.; and Siegel, R.~B. 2003.
\newblock The American civil rights tradition: Anticlassification or antisubordination.
\newblock \emph{Issues in Legal Scholarship}, 2(1).

\bibitem[{Barkan, Mersky, and Dunn(2009)}]{barkan2009fundamentals}
Barkan, S.~M.; Mersky, R.~M.; and Dunn, D.~J. 2009.
\newblock Fundamentals of legal research.

\bibitem[{Barocas, Hardt, and Narayanan(2023)}]{barocas2023fairness}
Barocas, S.; Hardt, M.; and Narayanan, A. 2023.
\newblock \emph{Fairness and machine learning: Limitations and opportunities}.
\newblock MIT Press.

\bibitem[{Bart, Teodorescu, and Morse(2024)}]{bart2024perceptions}
Bart, Y.; Teodorescu, M.~H.; and Morse, L. 2024.
\newblock Perceptions of algorithmic criteria: The role of procedural fairness.

\bibitem[{Basavaraju(2009)}]{basavaraju2009reservation}
Basavaraju, C. 2009.
\newblock Reservation under the constitution of India: Issues and Perspectives.
\newblock \emph{Journal of the Indian Law Institute}, 51(2): 267--274.

\bibitem[{Berk et~al.(2021)Berk, Heidari, Jabbari, Kearns, and Roth}]{berk2021fairness}
Berk, R.; Heidari, H.; Jabbari, S.; Kearns, M.; and Roth, A. 2021.
\newblock Fairness in criminal justice risk assessments: The state of the art.
\newblock \emph{Sociological Methods \& Research}, 50(1): 3--44.

\bibitem[{Bowen(2009)}]{bowen2009document}
Bowen, G.~A. 2009.
\newblock Document analysis as a qualitative research method.
\newblock \emph{Qualitative research journal}, 9(2): 27--40.

\bibitem[{{California Code of Regulations}(2010)}]{California_regulations}
{California Code of Regulations}. 2010.
\newblock California Static Risk Assessment.
\newblock ({Cal. Code Regs. tit. 15, {\S} 3768.1}).

\bibitem[{{California Penal Code}(2021)}]{California2021}
{California Penal Code}. 2021.
\newblock Pretrial risk assessment tools; legislative intent; definitions; validation; public information; report on outcomes and potential biases in pretrial release.
\newblock ({Cal.Penal Code {\S} 1320.35}).

\bibitem[{Calvi and Kotzinos(2023)}]{calvi2023enhancing}
Calvi, A.; and Kotzinos, D. 2023.
\newblock Enhancing AI fairness through impact assessment in the European Union: a legal and computer science perspective.
\newblock In \emph{Proceedings of the 2023 ACM conference on fairness, accountability, and transparency}, 1229--1245.

\bibitem[{Caudy, Durso, and Taxman(2013)}]{caudy2013well}
Caudy, M.~S.; Durso, J.~M.; and Taxman, F.~S. 2013.
\newblock How well do dynamic needs predict recidivism? Implications for risk assessment and risk reduction.
\newblock \emph{Journal of Criminal Justice}, 41(6): 458--466.

\bibitem[{Chouldechova(2017)}]{chouldechova2017fair}
Chouldechova, A. 2017.
\newblock Fair prediction with disparate impact: A study of bias in recidivism prediction instruments.
\newblock \emph{Big data}, 5(2): 153--163.

\bibitem[{Colquitt(2001)}]{colquitt2001dimensionality}
Colquitt, J.~A. 2001.
\newblock On the dimensionality of organizational justice: a construct validation of a measure.
\newblock \emph{Journal of applied psychology}, 86(3): 386.

\bibitem[{Comber and Rossitto(2023)}]{comber2023regulating}
Comber, R.; and Rossitto, C. 2023.
\newblock Regulating responsibility: Environmental sustainability, law, and the platformisation of waste management.
\newblock In \emph{Proceedings of CHI 2023}, 1--19.

\bibitem[{{Consolidated Laws of New York}(2011)}]{NewYork2011}
{Consolidated Laws of New York}. 2011.
\newblock Board of examiners of sex offenders.
\newblock ({N.Y. Correct. Law {\S} 168-l }).

\bibitem[{Corbett-Davies et~al.(2023)Corbett-Davies, Nilforoshan, Shroff, and Goel}]{corbett2023measure}
Corbett-Davies, S.; Nilforoshan, H.; Shroff, R.; and Goel, S. 2023.
\newblock The measure and mismeasure of fairness.
\newblock \emph{Journal of Machine Learning Research}.

\bibitem[{{Council of European Union}(2000{\natexlab{a}})}]{framework_equality_directive}
{Council of European Union}. 2000{\natexlab{a}}.
\newblock {Council Directive 2000/78/EC of 27 November 2000 establishing a general framework for equal treatment in employment and occupation.}
\newblock Official Journal L 303 (2000), 16–22.

\bibitem[{{Council of European Union}(2000{\natexlab{b}})}]{racial_equality_directive}
{Council of European Union}. 2000{\natexlab{b}}.
\newblock {Racial Equality Directive. Council Directive 2000/43/EC of 29 June 2000 implementing the principle of equal treatment between persons irrespective of racial or ethnic origin.}
\newblock Official Journal L 180 (2000), 22–26.

\bibitem[{{Council of European Union}(2004)}]{gender_equality_directive_2004}
{Council of European Union}. 2004.
\newblock {Council Directive 2004/113/EC of 13 December 2004 implementing the principle of equal treatment between men and women in the access to and supply of goods and services.}
\newblock Official Journal L 373 (2004), 37–43.

\bibitem[{{Council of European Union}(2006)}]{gender_equality_directive_2006}
{Council of European Union}. 2006.
\newblock {Directive 2006/54/EC of the European Parliament and of the Council of 5 July 2006 on the implementation of the principle of equal opportunities and equal treatment of men and women in matters of employment and occupation.}
\newblock Official Journal L 204 (2006), 23—-36.

\bibitem[{{Court of Appeals of Virginia}(2004)}]{2004brooks}
{Court of Appeals of Virginia}. 2004.
\newblock Brooks v. Commonwealth.
\newblock ({Record No. 2540-02-3 (Va. Ct. App. Jan. 28, 2004)}).

\bibitem[{Cui, Cheng, and Wiesner(2019)}]{cui2019judicial}
Cui, W.; Cheng, J.; and Wiesner, D. 2019.
\newblock Judicial review of government actions in China.
\newblock \emph{China Perspectives}, 2019(2019-1): 35--44.

\bibitem[{Cui(2020)}]{cui2020artificial}
Cui, Y. 2020.
\newblock \emph{Artificial intelligence and judicial modernization}.
\newblock Springer.

\bibitem[{Delgado(2020)}]{delgado2020sociotechnical}
Delgado, F.~A. 2020.
\newblock Sociotechnical Design in Legal Algorithmic Decision-Making.
\newblock In \emph{Companion Publication of CSCW 2020}, 111--115.

\bibitem[{{District Court, SD New York}(2021)}]{2021flores}
{District Court, SD New York}. 2021.
\newblock Flores v. Stanford.
\newblock ({No. 18CIV02468VBJCM, 2021 WL 4441614 (S.D.N.Y. Sept. 28, 2021)}).

\bibitem[{Dressel and Farid(2018)}]{dressel2018accuracy}
Dressel, J.; and Farid, H. 2018.
\newblock The accuracy, fairness, and limits of predicting recidivism.
\newblock \emph{Science advances}, 4(1).

\bibitem[{Dwork et~al.(2012)Dwork, Hardt, Pitassi, Reingold, and Zemel}]{dwork2012fairness}
Dwork, C.; Hardt, M.; Pitassi, T.; Reingold, O.; and Zemel, R. 2012.
\newblock Fairness through awareness.
\newblock In \emph{Proceedings of ITCS 2012}, 214--226.

\bibitem[{Esfahani et~al.(2024)Esfahani, De~Toni, Lepri, Passerini, Tentori, and Zancanaro}]{esfahani2024preference}
Esfahani, S.; De~Toni, G.; Lepri, B.; Passerini, A.; Tentori, K.; and Zancanaro, M. 2024.
\newblock Preference Elicitation in Interactive and User-centered Algorithmic Recourse: an Initial Exploration.
\newblock In \emph{Proceedings of the 32nd ACM Conference on User Modeling, Adaptation and Personalization}, 249--254.

\bibitem[{Goyal et~al.(2024)Goyal, Baumler, Nguyen, and Daum{\'e}~III}]{goyal2024impact}
Goyal, N.; Baumler, C.; Nguyen, T.; and Daum{\'e}~III, H. 2024.
\newblock The Impact of Explanations on Fairness in Human-AI Decision-Making: Protected vs Proxy Features.
\newblock In \emph{Proceedings of IUI 2024}, 155--180.

\bibitem[{Gray et~al.(2021)Gray, Santos, Bielova, Toth, and Clifford}]{gray2021dark}
Gray, C.~M.; Santos, C.; Bielova, N.; Toth, M.; and Clifford, D. 2021.
\newblock Dark patterns and the legal requirements of consent banners: An interaction criticism perspective.
\newblock In \emph{Proceedings of CHI 2021}, 1--18.

\bibitem[{Green and Chen(2019)}]{green2019disparate}
Green, B.; and Chen, Y. 2019.
\newblock Disparate interactions: An algorithm-in-the-loop analysis of fairness in risk assessments.
\newblock In \emph{Proceedings of FAccT (FAT) 2019}, 90--99.

\bibitem[{Grgi{\'c}-Hla{\v{c}}a et~al.(2018)Grgi{\'c}-Hla{\v{c}}a, Zafar, Gummadi, and Weller}]{grgic2018beyond}
Grgi{\'c}-Hla{\v{c}}a, N.; Zafar, M.~B.; Gummadi, K.~P.; and Weller, A. 2018.
\newblock Beyond distributive fairness in algorithmic decision making: Feature selection for procedurally fair learning.
\newblock In \emph{Proceedings of the AAAI conference on artificial intelligence}, volume~32.

\bibitem[{Grimpe, Hartswood, and Jirotka(2014)}]{grimpe2014towards}
Grimpe, B.; Hartswood, M.; and Jirotka, M. 2014.
\newblock Towards a closer dialogue between policy and practice: responsible design in HCI.
\newblock In \emph{Proceedings of CHI 2014}, 2965--2974.

\bibitem[{Grossman, Nyarko, and Goel(2024)}]{grossman2024reconciling}
Grossman, J.; Nyarko, J.; and Goel, S. 2024.
\newblock Reconciling Legal and Empirical Conceptions of Disparate Impact: An Analysis of Police Stops Across California.
\newblock \emph{Journal of Law and Empirical Analysis}.

\bibitem[{Hamilton et~al.(2016)Hamilton, Kigerl, Campagna, Barnoski, Lee, Van~Wormer, and Block}]{hamilton2016designed}
Hamilton, Z.; Kigerl, A.; Campagna, M.; Barnoski, R.; Lee, S.; Van~Wormer, J.; and Block, L. 2016.
\newblock Designed to fit: The development and validation of the STRONG-R recidivism risk assessment.
\newblock \emph{Criminal Justice and behavior}, 43(2): 230--263.

\bibitem[{Hanson and Morton-Bourgon(2009)}]{hanson2009accuracy}
Hanson, R.~K.; and Morton-Bourgon, K.~E. 2009.
\newblock The accuracy of recidivism risk assessments for sexual offenders: a meta-analysis of 118 prediction studies.
\newblock \emph{Psychological assessment}, 21(1): 1.

\bibitem[{Hu(2023)}]{hu2023what}
Hu, L. 2023.
\newblock What is ``Race'' in Algorithmic Discrimination on the Basis of Race?
\newblock \emph{Journal of Moral Philosophy}, 1--26.

\bibitem[{{Illinois Administrative Code}(2017)}]{Illinois_admin}
{Illinois Administrative Code}. 2017.
\newblock Adult Sex Offender Evaluation and Treatment - Standards of Practice - Risk Assessment.
\newblock ({Ill. Admin. Code tit. 20, {\S} 1905.60}).

\bibitem[{{Illinois Compiled Statutes}(2023)}]{Illinois2023}
{Illinois Compiled Statutes}. 2023.
\newblock Statewide risk-assessment tool.
\newblock ({\S} 725 ILCS 5/110-6.4).

\bibitem[{Imrey and Dawid(2015)}]{imrey2015commentary}
Imrey, P.~B.; and Dawid, A.~P. 2015.
\newblock A commentary on statistical assessment of violence recidivism risk.
\newblock \emph{Statistics and Public Policy}, 2(1): 1--18.

\bibitem[{Johnson(2021)}]{johnson2021algorithmic}
Johnson, G.~M. 2021.
\newblock Algorithmic bias: on the implicit biases of social technology.
\newblock \emph{Synthese}, 198(10): 9941--9961.

\bibitem[{Karimi, Sch{\"o}lkopf, and Valera(2021)}]{karimi2021algorithmic}
Karimi, A.-H.; Sch{\"o}lkopf, B.; and Valera, I. 2021.
\newblock Algorithmic recourse: from counterfactual explanations to interventions.
\newblock In \emph{Proceedings of FAccT 2021}, 353--362.

\bibitem[{Karusala et~al.(2024)Karusala, Upadhyay, Veeraraghavan, and Gajos}]{karusala2024understanding}
Karusala, N.; Upadhyay, S.; Veeraraghavan, R.; and Gajos, K.~Z. 2024.
\newblock Understanding Contestability on the Margins: Implications for the Design of Algorithmic Decision-making in Public Services.
\newblock In \emph{Proceedings of CHI 2024}, 1--16.

\bibitem[{Katsiyannis et~al.(2018)Katsiyannis, Whitford, Zhang, and Gage}]{katsiyannis2018adult}
Katsiyannis, A.; Whitford, D.~K.; Zhang, D.; and Gage, N.~A. 2018.
\newblock Adult recidivism in United States: A meta-analysis 1994--2015.
\newblock \emph{Journal of Child and Family Studies}, 27: 686--696.

\bibitem[{Kleinberg, Mullainathan, and Raghavan(2017)}]{kleinberg2017inherent}
Kleinberg, J.; Mullainathan, S.; and Raghavan, M. 2017.
\newblock {Inherent Trade-Offs in the Fair Determination of Risk Scores}.
\newblock In \emph{Proceedings of ITCS 2017}, volume~67, 43:1--43:23. Schloss Dagstuhl -- Leibniz-Zentrum f{\"u}r Informatik.

\bibitem[{Lagioia, Rovatti, and Sartor(2023)}]{lagioia2023algorithmic}
Lagioia, F.; Rovatti, R.; and Sartor, G. 2023.
\newblock Algorithmic fairness through group parities? The case of COMPAS-SAPMOC.
\newblock \emph{AI \& society}, 38(2): 459--478.

\bibitem[{Lee et~al.(2019)Lee, Jain, Cha, Ojha, and Kusbit}]{lee2019procedural}
Lee, M.~K.; Jain, A.; Cha, H.~J.; Ojha, S.; and Kusbit, D. 2019.
\newblock Procedural justice in algorithmic fairness: Leveraging transparency and outcome control for fair algorithmic mediation.
\newblock \emph{Proceedings of the ACM on Human-Computer Interaction}, 3(CSCW): 1--26.

\bibitem[{Linos and Carlson(2017)}]{linos2017qualitative}
Linos, K.; and Carlson, M. 2017.
\newblock Qualitative methods for law review writing.
\newblock \emph{U. Chi. L. Rev.}, 84: 213.

\bibitem[{Lyons, Velloso, and Miller(2021)}]{lyons2021conceptualising}
Lyons, H.; Velloso, E.; and Miller, T. 2021.
\newblock Conceptualising contestability: Perspectives on contesting algorithmic decisions.
\newblock \emph{Proceedings of the ACM on Human-Computer Interaction}, 5(CSCW1): 1--25.

\bibitem[{Mann, Hanson, and Thornton(2010)}]{mann2010assessing}
Mann, R.~E.; Hanson, R.~K.; and Thornton, D. 2010.
\newblock Assessing risk for sexual recidivism: Some proposals on the nature of psychologically meaningful risk factors.
\newblock \emph{Sexual Abuse}, 22(2): 191--217.

\bibitem[{Mayson(2019)}]{mayson2019bias}
Mayson, S.~G. 2019.
\newblock Bias in, bias out.
\newblock \emph{The Yale Law Journal}, 128(8): 2218--2300.

\bibitem[{{Michigan Court of Appeals}(2019)}]{2019people}
{Michigan Court of Appeals}. 2019.
\newblock People v. Younglove.
\newblock ({No. 341901 (Mich. Ct. App. Feb. 21, 2019)}).

\bibitem[{Nakao et~al.(2022)Nakao, Stumpf, Ahmed, Naseer, and Strappelli}]{nakao2022toward}
Nakao, Y.; Stumpf, S.; Ahmed, S.; Naseer, A.; and Strappelli, L. 2022.
\newblock Toward involving end-users in interactive human-in-the-loop AI fairness.
\newblock \emph{ACM Transactions on Interactive Intelligent Systems (TiiS)}, 12(3): 1--30.

\bibitem[{Narayanan et~al.(2024)Narayanan, Nagpal, McGuire, Schweitzer, and De~Cremer}]{narayanan2024fairness}
Narayanan, D.; Nagpal, M.; McGuire, J.; Schweitzer, S.; and De~Cremer, D. 2024.
\newblock Fairness perceptions of artificial intelligence: A review and path forward.
\newblock \emph{International Journal of Human--Computer Interaction}, 40(1): 4--23.

\bibitem[{{New Jersey Statutes}(2024)}]{NewJersey}
{New Jersey Statutes}. 2024.
\newblock Statewide Pretrial Services Program; establishment; risk assessment instrument; monitoring of eligible defendants on conditional release (Proposed Legislation).
\newblock ({N.J. Stat. {\S} 2A:162-25}).

\bibitem[{Nguyen et~al.(2025)Nguyen, Xu, Nguyen-Le, Lazar, Braman, Daum{\'e}~III, and Jelveh}]{nguyen2025demographic}
Nguyen, T.~T.; Xu, J.; Nguyen-Le, P.-A.; Lazar, J.; Braman, D.; Daum{\'e}~III, H.; and Jelveh, Z. 2025.
\newblock Which Demographic Features Are Relevant for Individual Fairness Evaluation of US Recidivism Risk Assessment Tools?
\newblock \emph{The 20th International Conference on Artificial Intelligence and Law. arXiv preprint arXiv:2505.09868}.

\bibitem[{Nishi(2019)}]{nishi2019privatizing}
Nishi, A. 2019.
\newblock Privatizing sentencing: A delegation framework for recidivism risk assessment.
\newblock \emph{Colum. L. Rev.}, 119: 1671.

\bibitem[{{Oregon Administrative Code}(2015{\natexlab{a}})}]{Oregon_admin_probation}
{Oregon Administrative Code}. 2015{\natexlab{a}}.
\newblock Case Management System (Community Corrections) - Risk Assessment.
\newblock ({Or. Admin. R. 291-078-0020}).

\bibitem[{{Oregon Administrative Code}(2015{\natexlab{b}})}]{Oregon_admin}
{Oregon Administrative Code}. 2015{\natexlab{b}}.
\newblock Sex Offender Risk Assessment Methodology.
\newblock ({Or. Admin. R. 255-085-0020}).

\bibitem[{Papagianneas and Junius(2023)}]{papagianneas2023fairness}
Papagianneas, S.; and Junius, N. 2023.
\newblock Fairness and justice through automation in China's smart courts.
\newblock \emph{Computer Law \& Security Review}, 51: 105897.

\bibitem[{Pedreshi, Ruggieri, and Turini(2008)}]{pedreshi2008discrimination}
Pedreshi, D.; Ruggieri, S.; and Turini, F. 2008.
\newblock Discrimination-aware data mining.
\newblock In \emph{Proceedings of SIGKDD 2008}, 560--568.

\bibitem[{{Pennsylvania Administrative Code}(2019{\natexlab{a}})}]{Pennsylvania_admin}
{Pennsylvania Administrative Code}. 2019{\natexlab{a}}.
\newblock Sentence Risk Assessment Instrument methodology.
\newblock ({204 Pa.Code {\S} 305.2}).

\bibitem[{{Pennsylvania Administrative Code}(2019{\natexlab{b}})}]{Pennsylvania2019}
{Pennsylvania Administrative Code}. 2019{\natexlab{b}}.
\newblock Sentence Risk Assessment Instrument methodology.
\newblock ({204 Pa. Stat. Ann. {\S} 305.2}).

\bibitem[{Pessach and Shmueli(2022)}]{pessach2022review}
Pessach, D.; and Shmueli, E. 2022.
\newblock A review on fairness in machine learning.
\newblock \emph{ACM Computing Surveys (CSUR)}, 55(3): 1--44.

\bibitem[{Phillips(2019)}]{phillips2019practical}
Phillips, M. 2019.
\newblock \emph{A Practical Guide to Legal Research and Analysis for Paralegal and Legal Studies Students (Higher Education Coursebook)}.
\newblock West Academic Publishing.

\bibitem[{Rigotti and Fosch-Villaronga(2024)}]{rigotti2024fairness}
Rigotti, C.; and Fosch-Villaronga, E. 2024.
\newblock Fairness, AI \& recruitment.
\newblock \emph{Computer Law \& Security Review}, 53: 105966.

\bibitem[{Rudin(2019)}]{rudin2019stop}
Rudin, C. 2019.
\newblock Stop explaining black box machine learning models for high stakes decisions and use interpretable models instead.
\newblock \emph{Nature machine intelligence}, 1(5): 206--215.

\bibitem[{Ryan, Nadal, and Doherty(2023)}]{ryan2023integrating}
Ryan, S.; Nadal, C.; and Doherty, G. 2023.
\newblock Integrating fairness in the software design process: An interview study with HCI and ML experts.
\newblock \emph{IEEE Access}, 11: 29296--29313.

\bibitem[{Samaha(2016)}]{samaha2016criminal}
Samaha, J. 2016.
\newblock \emph{Criminal law}.
\newblock Cengage Learning.

\bibitem[{Sambasivan et~al.(2021)Sambasivan, Arnesen, Hutchinson, Doshi, and Prabhakaran}]{sambasivan2021re}
Sambasivan, N.; Arnesen, E.; Hutchinson, B.; Doshi, T.; and Prabhakaran, V. 2021.
\newblock Re-imagining algorithmic fairness in india and beyond.
\newblock In \emph{Proceedings of the 2021 ACM conference on fairness, accountability, and transparency}, 315--328.

\bibitem[{Singh and Jackson(2021)}]{singh2021seeing}
Singh, R.; and Jackson, S. 2021.
\newblock Seeing like an infrastructure: Low-resolution citizens and the Aadhaar identification project.
\newblock \emph{Proceedings of the ACM on Human-Computer Interaction}, 5(CSCW2): 1--26.

\bibitem[{Sreenivasan et~al.(2000)Sreenivasan, Kirkish, Garrick, Weinberger, and Phenix}]{sreenivasan2000actuarial}
Sreenivasan, S.; Kirkish, P.; Garrick, T.; Weinberger, L.~E.; and Phenix, A. 2000.
\newblock Actuarial risk assessment models: A review of critical issues related to violence and sex-offender recidivism assessments.
\newblock \emph{Journal-American Academy of Psychiatry and the Law}, 28: 438--448.

\bibitem[{Stevenson and Doleac(2022)}]{stevenson2022algorithmic}
Stevenson, M.~T.; and Doleac, J.~L. 2022.
\newblock Algorithmic risk assessment in the hands of humans.
\newblock \emph{Available at SSRN 3489440}.

\bibitem[{{Supreme Court of India}(2021)}]{nitisha2021}
{Supreme Court of India}. 2021.
\newblock LT. Col Nitisha v. Union of India.

\bibitem[{{The European Parliament and Council}(2016)}]{gdpr_2016}
{The European Parliament and Council}. 2016.
\newblock {Regulation (EU) 2016/679 of the European Parliament and of the Council of 27 April 2016 on the protection of natural persons with regard to the processing of personal data and on the free movement of such data, and repealing Directive 95/46/EC (General Data Protection Regulation)}.
\newblock Official Journal L 119/1.

\bibitem[{{The European Parliament and Council}(2022)}]{dsa_2022}
{The European Parliament and Council}. 2022.
\newblock {Regulation (EU) 2022/2065 of the European Parliament and of the Council of 19 October 2022 on a Single Market For Digital Services and amending Directive 2000/31/EC (Digital Services Act)}.
\newblock Official Journal L 277/1.

\bibitem[{{The European Parliament and Council}(2024)}]{eu_ai_act_2024}
{The European Parliament and Council}. 2024.
\newblock {Artificial Intelligence Act}.
\newblock Official Journal of the European Union, L 2024/1689, 12 July 2024, pp.~1–144.

\bibitem[{Tobler(2005)}]{tobler2005indirect}
Tobler, C. 2005.
\newblock \emph{Indirect discrimination: a case study into the development of the legal concept of indirect discrimination under EC law}, volume~10.
\newblock Intersentia nv.

\bibitem[{{U.S. Supreme Court}(1920)}]{1920fs}
{U.S. Supreme Court}. 1920.
\newblock FS Royster Guano Co. v. Virginia.
\newblock ({253 U.S. 412}).

\bibitem[{{U.S. Supreme Court}(1944)}]{1944korematsu}
{U.S. Supreme Court}. 1944.
\newblock Korematsu v. United States.
\newblock (323 U.S. 214).

\bibitem[{{U.S. Supreme Court}(1971)}]{1971graham}
{U.S. Supreme Court}. 1971.
\newblock Graham v. Richardson.
\newblock (403 U.S. 365).

\bibitem[{{U.S. Supreme Court}(1976{\natexlab{a}})}]{1976hampton}
{U.S. Supreme Court}. 1976{\natexlab{a}}.
\newblock Hampton v. Mow Sun Wong.
\newblock (426 U.S. 88).

\bibitem[{{U.S. Supreme Court}(1976{\natexlab{b}})}]{1976washington}
{U.S. Supreme Court}. 1976{\natexlab{b}}.
\newblock Washington v. Davis.
\newblock ({426 U.S. 229}).

\bibitem[{{U.S. Supreme Court}(1977)}]{1977dobbert}
{U.S. Supreme Court}. 1977.
\newblock Dobbert v. Florida.
\newblock ({432 U.S. 282}).

\bibitem[{{U.S. Supreme Court}(1978)}]{1978foley}
{U.S. Supreme Court}. 1978.
\newblock Foley v. Connelie.
\newblock (435 U. S. 291).

\bibitem[{{U.S. Supreme Court}(1979)}]{1979ambach}
{U.S. Supreme Court}. 1979.
\newblock Ambach v. Norwick.
\newblock (441 U. S. 68).

\bibitem[{{U.S. Supreme Court}(1982)}]{1982cabell}
{U.S. Supreme Court}. 1982.
\newblock Cabell v. Chavez-Salido.
\newblock (454 U. S. 432).

\bibitem[{{U.S. Supreme Court}(1984)}]{1984bernal}
{U.S. Supreme Court}. 1984.
\newblock Bernal v. Fainter.
\newblock (467 U.S. 216 (1984)).

\bibitem[{{U.S. Supreme Court}(1985)}]{1985cleburne}
{U.S. Supreme Court}. 1985.
\newblock Cleburne v. Cleburne Living Center, Inc.
\newblock ({473 U.S. 432}).

\bibitem[{{U.S. Supreme Court}(1988)}]{1988clark}
{U.S. Supreme Court}. 1988.
\newblock Clark v. Jeter.
\newblock (473 U.S. 432).

\bibitem[{{U.S. Supreme Court}(1995)}]{1995adarand}
{U.S. Supreme Court}. 1995.
\newblock Adarand Constructors, Inc. v. Peña.
\newblock (515 U.S. 200).

\bibitem[{{U.S. Supreme Court}(1996)}]{1996us}
{U.S. Supreme Court}. 1996.
\newblock U.S. v. Armstrong.
\newblock ({517 U.S. 456}).

\bibitem[{Vaccaro et~al.(2019)Vaccaro, Karahalios, Mulligan, Kluttz, and Hirsch}]{vaccaro2019contestability}
Vaccaro, K.; Karahalios, K.; Mulligan, D.~K.; Kluttz, D.; and Hirsch, T. 2019.
\newblock Contestability in algorithmic systems.
\newblock In \emph{Companion Publication CSCW 2019}, 523--527.

\bibitem[{Washington(2018)}]{washington2018argue}
Washington, A.~L. 2018.
\newblock How to argue with an algorithm: Lessons from the COMPAS-ProPublica debate.
\newblock \emph{Colo. Tech. LJ}, 17: 131.

\bibitem[{Weerts et~al.(2023)Weerts, Xenidis, Tarissan, Olsen, and Pechenizkiy}]{weerts2023algorithmic}
Weerts, H.; Xenidis, R.; Tarissan, F.; Olsen, H.~P.; and Pechenizkiy, M. 2023.
\newblock Algorithmic unfairness through the lens of EU non-discrimination law: Or why the law is not a decision tree.
\newblock In \emph{Proceedings of the 2023 ACM Conference on Fairness, Accountability, and Transparency}, 805--816.

\bibitem[{Wexler(2018)}]{wexler2018life}
Wexler, R. 2018.
\newblock Life, liberty, and trade secrets: Intellectual property in the criminal justice system.
\newblock \emph{Stan. L. Rev.}, 70: 1343.

\bibitem[{{Wisconsin Supreme Court}(2016)}]{2016state}
{Wisconsin Supreme Court}. 2016.
\newblock State v. Loomis.
\newblock ({2016 WI 68, 371 Wis. 2d 235, 881 N.W.2d 749}).

\bibitem[{Xiang and Raji(2019)}]{xiang2019legal}
Xiang, A.; and Raji, I.~D. 2019.
\newblock On the legal compatibility of fairness definitions.
\newblock \emph{Workshop on Human-Centric Machine Learning at the 33rd Conference on Neural Information Processing Systems (NeurIPS 2019)}.

\bibitem[{Yacoby et~al.(2022)Yacoby, Green, Griffin~Jr, and Doshi-Velez}]{yacoby2022if}
Yacoby, Y.; Green, B.; Griffin~Jr, C.~L.; and Doshi-Velez, F. 2022.
\newblock “If it didn’t happen, why would I change my decision?”: How Judges Respond to Counterfactual Explanations for the Public Safety Assessment.
\newblock In \emph{Proceedings of HCOMP 2024}, volume~10, 219--230.

\bibitem[{Yang and Dobbie(2020)}]{yang2020equal}
Yang, C.~S.; and Dobbie, W. 2020.
\newblock Equal protection under algorithms: A new statistical and legal framework.
\newblock \emph{Mich. L. Rev.}, 119: 291.

\bibitem[{Yfantidou et~al.(2023)Yfantidou, Constantinides, Spathis, Vakali, Quercia, and Kawsar}]{yfantidou2023state}
Yfantidou, S.; Constantinides, M.; Spathis, D.; Vakali, A.; Quercia, D.; and Kawsar, F. 2023.
\newblock The state of algorithmic fairness in mobile human-computer interaction.
\newblock In \emph{Proceedings of MobileHCI 2023}, 1--7.

\end{thebibliography}

\appendix

\section{Appendix: Detailed Review of Fairness-related Sources of U.S. Law}
\label{appendix-law-review}
Before synthesizing the legal findings presented in this paper, we conducted a detailed review of the four main sources of federal and state laws: the U.S. Constitution, statutes, case laws, and regulations. In each category (source of law), we identify legal items that are either related to AI, recidivism risk assessment, or general fairness-related provisions (“equal protection” or “due process”), give a quick summary and analysis of each source, and quote the most relevant excerpt from the source, as presented in the 30-pages attachment below. Please note that not all legal items below go into the main narrative of the paper.

\clearpage

%\includepdf[pages=-,pagecommand={},width=\textwidth]{NIJ_Fairness_Thresholds_Law_Review_Appendix.pdf}



\section*{Supplementary material}

# Review of RRA and Fairness-related Sources of U.S. Law

# The U.S. Constitution

**1/ Eighth Amendment**
Relevant Excerpt: "Excessive bail shall not be required…"
**Analysis of the 8th Amendment**: Since the amount of bail may increase with higher predicted risk of recidivism by an AI model, this Amendment shows that the accuracy and fairness of those risk assessment models might be framed as a constitutional issue.


**2/ Fifth Amendment and Fourteenth Amendment**
**5th Amendment** (scale: federal courts):
Relevant Excerpt: "No person shall be [...] deprived of life, **liberty**, or property, without **due process** of law."

**14th Amendment** (scale: state courts):
Relevant Excerpt: "...nor shall any State deprive any person of life, **liberty**, or property, without **due process** of law; nor deny to any **person** within its jurisdiction the **equal protection** of the laws."

**Analysis of the 5th and 14th Amendment**: The discussion of 'liberty' is directly related to bail. While "due process" (applicable to both state and federal courts) is clearly procedural fairness, "equal protection" (applicable to only state courts as it does not show up in the 5th Amendment) can be interpreted as outcome-based fairness, which include group and individual fairness. Individual fairness, rather than group fairness, seems to better represent the original meaning of "equal protection". One counterexample is that if only group fairness (and not individual fairness) is guaranteed by equal protection clause, then a random rule of assigning 50% recidivism rate to everyone, which satisfies group fairness but not individual fairness, will violate "equal protection". However, with the development of case laws, anti-discrimination based on protected group membership (group fairness) becomes an extended meaning of equal protection. One law review article that discusses the extension from "equal protection" to group-based anti-discrimination is Jack M. Balkin & Reva B. Siegel, "The American Civil Rights Tradition: Anticlassification or Antisubordination?", 58 U. MIAMI L. REV. 9, 10 (2003).

# Case Law (Judicial)

**1/ State v. Loomis, 2016 WI 68, 371 Wis. 2d 235, 881 N.W.2d 749**

This case concerns a defendant who was suspected to be the driver in a drive-by shooting and in a plea bargain pleaded guilty to two less serious charges, i.e. "Attempting to flee or elude a traffic officer (PTAC)" and "Operating a motor vehicle without the owner's consent", in return for the three more serious charges to be not pressed, but still read-in for the sentencing judge to consider as potential aggravating factors in determining the sentence (to the extent that the total number of prison years does not exceed the upper bound for the two plead-guilty charges combined). The defendant also argues that the sentencing judge's consideration of COMPAS recidivism risk score violates his constitutional right to "due process", on three grounds: firstly, COMPAS is proprietary (thus its algorithm cannot be investigated for accuracy); secondly, it violates the defendant's right to an "individualized sentence"; finally, it takes gender into account for its prediction. With those arguments, the defendant requests an alternative hearing for sentencing.

On the read-in issue, the Wisconsin Supreme Court rules that the sentencing judge followed a common and legitimate legal practice. Most related to our law review work, on the first two due process issue with respect to COMPAS use, this court rules that COMPAS is still useful enough and has been validated by several other states (not yet Wisconsin at the decision time) that future courts should still consider COMPAS risk score, only with explicit caution that COMPAS accuracy has not been fully validated yet and the tool only assess risk at the group level, not the individual level. On the third due process issue with gender as a feature, this court advocates for the use of this feature on the ground that men and women have been historically shown to have different recidivism rates.

"Ultimately, we conclude that if used properly, observing the limitations and cautions set forth herein, a circuit court's consideration of a COMPAS risk assessment at sentencing does not violate a defendant's right to due process […] Specifically, any PSI containing a COMPAS risk assessment must inform the sentencing court about the following cautions regarding a COMPAS risk assessment's accuracy: (1) the proprietary nature of COMPAS has been invoked to prevent disclosure of information relating to how factors are weighed or how risk scores are to be determined; (2) risk assessment compares defendants to a national sample, but no cross-validation study for a Wisconsin population has yet been completed; (3) some studies of COMPAS risk assessment scores have raised questions about whether they disproportionately classify minority offenders as having a higher risk of recidivism; and (4) risk assessment tools must be constantly monitored and re-normed for accuracy due to changing populations and subpopulations [...] We determine that COMPAS's use of gender promotes accuracy that ultimately inures to the benefit of the justice system including defendants. Additionally, we determine that the defendant failed to meet his burden of showing that the sentencing court actually relied on gender as a factor in sentencing."

**Stage of criminal procedure**: highest-level state appellate court (state court of last resort)

**Binding case law?** Yes, on all trial and mid-level appellate courts in Wisconsin

**Is the case still considered good law?** Probably Yes. However, there are two WestLaw yellow flags (Negative Treatment) from the following cases:

**State v. Bolstad, 2021 WI App 81, 399 Wis. 2d 815, 967 N.W.2d 164.**
This case is decided by the Court of Appeals of Wisconsin (mid-level state appellate court), which does not have the authority to override the Wisconsin Supreme Court. The case outcome here is almost the opposite of State v. Loomis case outcome: the Court of Appeals here reversed the trial's sentencing decision because the trial court failed to mention the grounds of their sentence, especially their lack of explicit consideration for three factors listed in State v. Loomis: "the gravity of the offense, the character and rehabilitative needs of the defendant, and the need to protect the public". However, this decision adopts a similar line of reasoning compared to the Wisconsin Supreme Court, and the outcome difference just stems from the difference in key procedural facts from the lower sentencing courts.

**Henderson v. Stensberg, No. 18-CV-555-JDP, 2020 WL 1320820 (W.D. Wis. Mar. 20, 2020).**
This case is decided by the United States District Court, W.D. Wisconsin., which does not have the authority to override the Wisconsin Supreme Court due to the difference in jurisdictions (federal vs. state courts). This federal district court allows a claim of COMPAS racial bias to move forward on the defendant's ground of the "Equal Protection" clause in the Fourteenth Amendment. However, this decision does not necessarily contradict State v. Loomis, because the Wisconsin Supreme Court in State v. Loomis explicitly mention that they only consider refute the defendant's claim of COMPAS bias in terms of the "Due Process" clause, and not the "Equal Protection" clause because Loomis never claims the latter: "Notably, however, Loomis does not bring an *equal protection* challenge in this case. Thus, we address whether Loomis's constitutional *due process* right not to be sentenced on the basis of gender is violated if a circuit court considers a COMPAS risk assessment at sentencing."

**2/ Clark v. Jeter, 486 U.S. 456, 108 S. Ct. 1910, 100 L. Ed. 2d 465 (1988)**
The plaintiff in this case is a single mother who only decides to claim child support from the father ten years after their illegitimate child was born. All three levels of state court, i.e. trial court (Allegheny County Court of Common Pleas), mid-level appellate court (the Superior Court of Pennsylvania), and the Pennsylvania Supreme Court, have denied the mother's child support claim on the ground that Pennsylvania has a 6-year statute of limitation for child support claim on behalf of an illegitimate child (not applicable to children whose parents were married). According to those state courts, this 6-year statute of limitation was only replaced by a new 18-year statute of limitation after the point that the plaintiff's child had reached an age higher than 6 years old and therefore the 6-year statute of limitation was still the binding law for this case. The US Supreme Court does not address the question whether the 18-year statute can be applied retroactively since it is not properly raised to the state courts during earlier stages. Instead, the US Supreme Court rules that the old 6-year statute of limitation was unconstitutional in the first place because this policy discriminates among children based on illegitimacy status, a feature that according to federal "equal protection" case laws deserves "intermediate scrutiny". Taking into account this scrutiny standard, the US Supreme Court develops a two-pronged test for cases related to statutes of limitations in child support claims: firstly, the period must be long enough for the parent with custody to claim support; secondly, the limitation must be justifiable by reason of avoiding fraud. Considering the fact of this case, the US Supreme Court remains inconclusive about the first prong of the test, but they assert that the second prong has not been satisfied by the Pennsylvania government:

"In light of this authority, we conclude that Pennsylvania's 6–year statute of limitations violates the Equal Protection Clause. Even six years does not necessarily provide a reasonable opportunity to assert a claim on behalf of an illegitimate child [...]
We do not rest our decision on this ground, however, for it is not entirely evident that six years would necessarily be an unreasonable limitations period for child support actions involving illegitimate children. We are, however, confident that the 6–year statute of limitations is not substantially related to Pennsylvania's interest in avoiding the litigation of stale or fraudulent claims. In a number of circumstances, Pennsylvania permits the issue of paternity to be litigated more than six years after the birth of an illegitimate child."

The US Supreme Court determines that Pennsylvania's 6-year statute of limitation fails the intermediate scrutiny test. Therefore, the court rules in favor of the single mother.

Related to our law review work, although this case is a civil and not a criminal case, it sheds light on which types of features are fair or unfair for the government to discriminate based on, which might be applied (with caution) in the recidivism risk assessment context to determine the appropriate features for an AI model to use as input for its prediction. In particular, this decision recognizes the hierarchy of scrutiny for features in order of decreasing level of warranted scrutiny derived from the constitutional "Equal Protection" clause and further interpreted by previous case laws: strict scrutiny (e.g. race, national origin), intermediate or 'heightened' scrutiny (e.g. gender, illegitimacy or out-of-wedlock status), and rational basis (e.g. age).

**Stage of civil procedure**: highest-level federal appellate court (federal court of last resort)

**Binding case law?** Yes, on all federal US district and circuit courts, and all state courts to the extent that this decision does not conflict with state laws in areas preserved for state laws.

**Is the case still considered good law?** Probably Yes. However, there are 14 yellow flags on WestLaw. The most negative treatment of this case is the following series of cases brought before the California Supreme Court.

**In re Marriage Cases, 43 Cal. 4th 757, 183 P.3d 384 (2008)**

In this combined case, the California Supreme Court rules that with respect to the "equal protection" concept in the Constitution of California, the parallel level of scrutiny that the sexual orientation feature warrants is a state-level "strict scrutiny" standard which is comparable to the federal "strict scrutiny" standard. Their rationale is that sexual orientation deserves no less scrutiny than sex or gender, which according to previous California case laws belongs to the "strict scrutiny" rather than the "intermediate scrutiny" category (where federal cases like Clark v. Jeter put the sex or gender feature in): "Unlike decisions applying the federal equal protection clause, California cases continue to review, under strict scrutiny rather than intermediate scrutiny, those statutes that impose differential treatment on the basis of sex or gender."

Although this California Supreme Court's decision differs from the US Supreme Court on the hierarchy category for sex or gender, it is consistent with the state vs. federal law principle in which federal laws serve as a lower bound for individual rights and states are totally within their power to grant their people more but not fewer rights. By requiring more scrutiny on government's discrimination on the basis of sex or gender, California clearly grants their people more rights than the federal baseline. This higher scrutiny standard at the state level does not invalidate the federal hierarchy of scrutiny, so Clark v. Jeter is still considered good law.

**3/ Flores v. Stanford, No. 18CIV02468VBJCM, 2021 WL 4441614 (S.D.N.Y. Sept. 28, 2021)**

This decision in 2021 is just part of an ongoing civil case taken by the US district court for the Southern District of New York since 2019. The Plaintiffs are a group of life sentence inmates (led by Flores) who committed serious offenses during their juvenile years and the defendant is the New York State Board of Parole (represented by Stanford). The main claims by the Plaintiffs in this is that the parole board denies them meaningful parole chances due to the recidivism risk assessment scores by COMPAS, which according to the Plaintiffs unfairly penalize them based on their young age-at-offense-time feature. Regarding this decision in 2021 specifically, the court deals with a third party's request to prevent the Plaintiffs' expert witness (Dr. Rudin) from getting access to the algorithm behind COMPAS to determine whether the way this algorithm uses age of offense as a feature will violate the Plaintiffs' constitutional rights (e.g. it may if the algorithm gives an exceedingly high weight to this feature, especially if against young offenders). This third party is Northpointe (the company that develops COMPAS). Its ground for stopping the expert witness's access is that Dr. Rudin is an open critique of COMPAS and therefore disclosing the algorithm to her may jeopardize the company's existence. The court rejects this hypothesis by Northpointe by reaffirming that the stringent confidentiality measures which Dr. Rudin will need to comply with to get access to the algorithm are sufficient to address Northpointe's concerns while still affording the Plaintiffs a chance to substantiate their age-based constitutional claims: "Northpointe's motion for a protective order is GRANTED in part, and DENIED in part. Consistent with this Opinion & Order, the parties and Northpointe are directed to submit for the Court's review a supplemental protective order specifically governing Dr. Rudin's review of the Compelled Materials."

For the full context, an earlier decision in 2019 rejects the Plaintiff's Sixth Amendment claim (that the parole board's continuous denial of the parole option for the Plaintiffs amounts to a de facto sentence equivalent to "life without the possibility of parole") but moves their Eighth and Fourteenth Amendment claims (about individualized parole assessment) forward. A later decision in 2022 is just a procedural next step by the court to accept the Plaintiffs' motion asking the Defendant to "produce certain victim impact and community opposition statements".

With respect to our law review research, in the same jurisdiction of New York, while Section 168-l of the Consolidated Laws of New York explicitly requires sex offense recidivism risk assessment to take into account "the age of the sex offender at the time of the commission of the first sex offense" (noting that sex offense is still often considered a less serious offense than murder - the type of offense committed during juvenile years by the Flores and most of his co-Plaintiffs) this Flores v. Stanford (2021) decision indicates an implicit stance against the use of age in recidivism risk assessment by allowing expert inspection of the data used to train COMPAS to "help Plaintiffs substantiate their allegations that COMPAS punishes juvenile offenders for their youth, such that Defendants' reliance on this tool is constitutionally problematic" as the Court believes that "the weight COMPAS affords to youth in determining recidivism risk will be highly relevant to the underlying constitutional claims".

**Stage of civil procedure**: lowest-level (district) federal court

**Binding case law?** No (this is at the lowest-level federal court, so the decision only has persuasive authority for later cases)

**Is the case still considered good law?** Yes (no flags from WestLaw)

**4/ F.S. Royster Guano Co. v. Commonwealth of Virginia, 253 U.S. 412, 40 S. Ct. 560, 64 L. Ed. 989 (1920)**

The Plaintiff of this case (F. S. Royster Guano Company) is based in Virginia but has several businesses (fertilizer manufacturing plants) not only in Virginia but also other states. They were taxed by the Virginia government for the profits they make not only by their Virginia-based plants but also their businesses in other states. All previous state-level court decisions (including the decision by the highest state appellate court back then - the Supreme Court of Appeals of Virginia) have ruled in favor of the defendant. However, the US Supreme Court rules for the plaintiff on the ground that double income taxation, i.e. taxing a company for the same income (or profit) in a state outside Virginia by both Virginia and that other state, violates the company's right to "equal protection of the law" under the Fourteenth Amendment. The legal theory behind this presumed equal protection violation is that it is unfair for a Virginia-based company whose income in both Virginia and other states to be subject to double income taxation while a Virginia-based company whose entire income coming from outside Virginia is exempt from Virginia income tax.

"But this was not retrospective, and, for the reasons given, we are constrained to hold that so far as chapter 472 of the Laws of 1916 operated to impose upon plaintiff in error a tax upon income derived from business transacted and property located without the state because of the mere circumstance that it also derived income from business transacted and property located within the state, while at the same time, under chapter 495, other corporations deriving their existence and powers from the laws of the same state, and receiving income from business transacted and property located without the state, but none from sources within the state, were exempted from income taxes, there was an arbitrary discrimination amounting to a denial to plaintiff in error of the equal protection of the laws within the meaning of the Fourteenth Amendment."

With respect to our law review research, this case has a different context from most of our other cases: it deals with civil law concerning tax issues for corporations, not criminal law and individuals). However, this decision interprets the US Constitution-based "equal protection" clause in a direction highly consistent with today's individual fairness concept in the AI community (similar individuals should be treated the same) if we define 'similarly situated' as having a high similarity score from the distance function: "The Equal Protection Clause of the Fourteenth Amendment commands [...] essentially a direction that all persons similarly situated should be treated alike."

**Stage of civil procedure**: highest-level federal appellate court (court of last resort)

**Binding case law?** Yes, on all federal US district and circuit courts, and all state courts to the extent that this decision does not conflict with state laws in areas preserved for state laws.

**Is the case still considered good law?** Probably Yes. However, there are four yellow flags from WestLaw. The "Most Negative" treatment according to WestLaw comes from the following case:

Williams v. Gentry, No. 204CV01620KJDEJY, 2020 WL 3302971 (D. Nev. June 18, 2020)

This decision by the US District Court, D. Nevada. does not disregard any substantive arguments from F.S. Royster Guano Co. v. Commonwealth of Virginia, but simply criticizes this US Supreme Court decision as belonging to a historical period of too much judicial activism that this court believes to be interfering with the authority of the legislature. However, for our purpose, this 1920 US Supreme Court decision still functions within the court's presumed duty to interpret the constitution (what "Equal Protection" in the Fourteenth Amendment means). Furthermore, a federal district court decision does not have the authority to overthrow a federal highest-level appellate court decision.

**5/ People v. Younglove, No. 341901, 2019 WL 846117 (Mich. Ct. App. Feb. 21, 2019)**

This case is decided by the Michigan Court of Appeals (the intermediate-level appellate court in Michigan). Four criminal defendants (including Younglove) have entered pleas of guilty or no-contest, but they appealed their sentences on the ground that the presentence investigation reports (PSIR), which their respective sentencing judges are required to consider, include AI-based (i.e. COMPAS-generated) risk scores in several categories: "residential instability, criminal personality, substance abuse, social isolation, and criminal associates or peers". The defendants appeal their sentences to the Michigan Court of Appeals with a legal theory that the presumed use of COMPAS risk scores by sentencing judges violate their rights to "due process" and "individualized sentencing". However, the Michigan Court of Appeals disagrees with the defendants and affirmed the sentences because the court holds that the burden of proof (to show that the sentencing judge over-relies on COMPAS score) is on the defendants and that burden of proof has not been satisfied:

"Courts must consider the PSIR, but whether or how heavily to weigh the information contained therein remains within the court's discretion … defendants offer no evidence that their sentencing courts actually placed significant (or any) weight on the COMPAS assessments in crafting their sentences. Defendants have failed to carry their burden of showing that the inclusion of the information affected their substantial rights."

With respect to our law review research, the holding raises a potentially unfair burden of proof standard because a defendant has virtually no access to a sentencing judge's mental model to determine how much a judge relies on the AI risk score and whether that reliance is motivated by implicit bias against a protected feature of the defendant (e.g. race, national origin), which if so will violate the defendant's right to due process.

**Stage of criminal procedure**: mid-level state appellate court (previous stages where the sentencing of four defendants happen: Ingham Circuit Court,, Berrien Circuit Court, Ionia Circuit Court, and Missaukee Circuit Court)

**Binding case law?** Yes, for all trial courts in Michigan

**Is the case still considered good law?** Yes (no WestLaw flags)

**6/ Washington v. Davis, 426 U.S. 229, 96 S. Ct. 2040, 48 L. Ed. 2d 597 (1976)**

The Plaintiffs in this class action are black candidates for police recruitment in the District of Columbia who failed a verbal test as part of the recruitment process. They filed suits against the DC Police Department with the legal theory that the verbal test has racial discrimination, violating their constitutional "due process" and "equal protection" rights because the pass rate among black applicants is lower than the pass rate among white applicants. While a federal district court rules for the defendant, a federal circuit court rules for the plaintiff. The US Supreme Court ultimately rules for the defendant on the grounds that disparate or non-equalized pass rate (or more generally, positive output rate) across races is not sufficient proof for racial discrimination. Related to our law review research, this court's argument might hint that group fairness ('disproportionate impact') matters but there should be another school of fairness under the "equal protection" clause:

"We have not held that a law [...] is invalid under the Equal Protection Clause simply because it may affect a greater proportion of one race than of another. Disproportionate impact is not irrelevant, but it is not the sole touchstone of an invidious racial discrimination forbidden by the Constitution."

**Stage of civil procedure**: highest-level federal appellate court (court of last resort)

**Binding case law?** Yes, on all federal US district and circuit courts, and all state courts to the extent that this decision does not conflict with state laws in areas preserved for state laws.

**Is the case still considered good law?** Probably Yes. However, there are 30 yellow flags from WestLaw. The "Most Negative" treatment according to WestLaw comes from the following case:

Veasey v. Perry, 29 F. Supp. 3d 896 (S.D. Tex. 2014)

This case is brought by voters, voting right organizations, and some government bodies against a voting registration requirement in Texas that requires ID with photos, which black voters are less likely to have access to than white voters, on the grounds that this seemingly race-neutral policy has disparate impact and therefore discriminates against black voters. However, when we look at the references to Washington v. Davis in this case, the US District Court, S.D. Texas, Corpus Christi Division only cites Washington v. Davis as part of their summary of the plaintiffs' argument (disparate impact is still a relevant but not sole factor in determination of racial discrimination) and defendant's argument (that race-and color-neutral laws do not constitute racial discrimination). Therefore, this case does not argue against Washington v. Davis, nor does this federal district court have the authority to do so.

**7/ City of Cleburne, Tex. v. Cleburne Living Ctr., 473 U.S. 432, 105 S. Ct. 3249, 87 L. Ed. 2d 313 (1985)**

The Plaintiff in this case is Respondent Cleburne Living Center, Inc. (CLC), a proposed operator for a group home dedicated to serving mentally retarded persons, who brought suits against their city (Cleburne)'s government in Texas for requiring (and subsequently rejecting to grant the plaintiff) a special zoning permit which is required by a zoning ordinance. The US Supreme Court rules that mental retardation is a feature in the rational basis scrutiny category, so even though the bar to tolerate government's discriminatory policy with respect to this feature is low, the government must be able to prove a legitimate economic or social interest. However, the court determines that the City of Cleburne has failed to show any economic or social interest rather than their own prejudice against the mentally retarded:

"The record does not reveal any rational basis for believing that the proposed group home would pose any special threat to the city's legitimate interests. Requiring the permit in this case appears to rest on an irrational prejudice against the mentally retarded, including those who would occupy the proposed group home and who would live under the closely supervised and highly regulated conditions expressly provided for by state and federal law."

Therefore, the court rules for the Plaintiff and thereby waives the special zoning permit requirement for them.

Related to our law review research, a concise synergy of the relevance of the "equal protection" clause to both group and individual fairness is provided in this case as the US Supreme Court rules that "Discrimination, in the Fourteenth Amendment sense, connotes a substantive constitutional judgment that two individuals or groups are entitled to be treated equally with respect to something."

**Stage of civil procedure**: highest-level federal appellate court (court of last resort)

**Binding case law?** Yes, on all federal US district and circuit courts, and all state courts to the extent that this decision does not conflict with state laws in areas preserved for state laws.

**Is the case still considered good law?** Probably Yes. However, there are 138 yellow flags from WestLaw. The "Most Negative" treatment according to WestLaw comes from the following case:

Wolinski v. Abdulgader, No. 2:21-CV-02078-CKD P, 2022 WL 256483 (E.D. Cal. Jan. 27, 2022)

The Plaintiff in this case is a disabled inmate who brought a civil suit against his doctor in prison for denying him access to medical treatment and prescription. The US District Court, E.D. California here only cites the City of Cleburne, Tex. v. Cleburne Living Ctr decision as a reference for the standards of proof that plaintiff has to satisfy for an "Equal Protection" claim, either by showing that the doctor denies him medical access due to his membership in a protected group (in the direction of "group fairness") or that he is treated differently from a "similarly situated" individual. Therefore, this case does not argue against City of Cleburne, Tex. v. Cleburne Living Ctr, nor does this federal district court have the authority to do so.

**8/ Brooks v. Commonwealth. Court of Appeals of Virginia. No. 2540-02-3 (2004).**

This decision concerns an appeal by Brooks who was convicted in a trial of statutory rape. Brooks only appeals the sentencing judge's consideration of the sentencing guidelines recommendation, which has been updated with a longer sentence after the recommendation takes into account AI-based recidivism risk score. The Court of Appeals of Virginia rules for the Commonwealth on the grounds that the trial judge has full discretion in considering the sentencing guidelines recommendation or not and in setting the exact length of the sentence (as long as it still falls within the legislative limits, which is between 2 and 10 years for this case).

Related to our law review research, one legal problem from this case is that this Human-AI interaction element becomes an automatic shield to protect the judge's decisions against accusations of (individual/group) fairness violation in future appeals, even though there is no at- tempt to measure the explicit reliance function of judges on the risk assessment score. For example, in Brooks v. Commonwealth, even though there is evidence that the trial judge favors the AI risk assessment-based sentencing recommendation and dismisses the shorter active sentence in the non-AI recommendation, the Court of Appeals of Virginia rules that "the trial court properly exercised its discretion" without inquiring the level of reliance by the trial court on the AI risk assessment

**Stage of criminal procedure**: mid-level state appellate court

**Binding case law?** Yes, on all trial courts in Virginia

# Statutes (Legislative)

1/ 18 U.S.C. § 3142

**Bail Reform Act of 1984.**
**Release or detention of a defendant pending trial (Effective: since 1984)**

The Bail Reform Act of 1984 has been codified into many sections of Title 18 of the US Code (from § 3141 to § 3150, and § 3162), but the most substantive section that is relevant to our fairness in recidivism risk assessment research is § 3142 Release or detention of a defendant pending trial. The scope of its applicability, as noted in 18 U.S.C. § 3150, is only for federal criminal cases and therefore not binding for state courts. It outlines the two main objectives federal court officials should optimize for in pre-trial detention decisions: trial appearance and community safety, motivating the development of pre-trial AI models that predict either failure to appear or recidivism risk. Subsection § 3142(g) also lists defendants' "factors to be considered" by the judicial officials: current charge, evidence, personal history, characteristics (e.g. previous arrest, probation, or parole), and sources of collateral if granted bail. These factors might justify or invalidate the relevance of each feature of the defendant that an AI model collects to make predictions. Interestingly, demographic features which are often used in AI recidivism risk assessment tools in many states, e.g. age, gender, are not listed in this federal statute, even though § 3142(g) does not explicitly mention whether its list of factors is exhaustive.

One small caveat is that a federal district court decision U.S. v. Karper (847 F.Supp.2d 350) rules that a small part of this statute violates the Eighth Amendment against excessive bail. In particular, § 3142(c)(1)(b), which as a result of the Adam Walsh Act was incorporated into this statute as an amendment, requires electronic monitoring and many stringent restrictions for defendants on bail who are charged with crimes related to minors, one of which is child pornography receipt and possession as detailed in 18 U.S.C. § 2252A(a)(2). The court rules that the charge of child pornography receipt and possession alone is not sufficient to automatically condition the defendant's bail on electronic monitoring and other freedom of movement restrictions laid out in § 3142(c)(1)(b), but his flight risk, criminal records and other relevant features as stipulated in this statute must be considered. Although the decision does not have mandatory authority (not a binding precedent), it is a case study where **individual fairness** is emphasized: bail decision must holistically consider **all relevant features** of a defendant, not just the current charge.

Currently, there is a circuit split on whether § 3142(c)(1)(b) as a result of the Adam Walsh Act still holds. On the one hand, U.S. v. Karper (a district-court-level decision from New York, part of the Second Circuit), finds the Adam Walsh Act and thus 3142(c)(1)(b) unconstitutional. On the other hand, US v. Stephens, 594 F.3d 1033 (8th Cir.2010) and US v. Peeples, 630 F.3d 1136 (9th Cir.2010) from the Eighth and Ninth Circuits rule that the Adam Walsh Act is not necessarily unconstitutional as they believe the government's interest in preventing risks against minors may outweigh individual's liberty in the bail context. A more nuanced discussion of this circuit split can be found in the law review article 54 A.L.R. Fed. 2d 195, "Propriety of Pretrial Release and Bail Under Adam Walsh Child Protection and Safety Act" by Fern L. Kletter, J.D. (2011).

**Relevant excerpts from the statute:**

"§ 3142(g) Factors To Be Considered.—The judicial officer shall, in determining whether there are conditions of release that will reasonably assure the **appearance** of the person as required and the **safety** of any other person and the community, take into account the available information concerning—

    (1)the nature and circumstances of the **offense charged**, including whether the offense is a crime of violence, a violation of section 1591, a Federal crime of terrorism, or involves a minor victim or a controlled substance, firearm, explosive, or destructive device;

    (2)the weight of the **evidence** against the person;

    (3)the **history and characteristics of the person**, including—

        (A)the person's character, physical and mental condition, family ties, employment, financial resources, length of residence in the community, community ties, past conduct, history relating to drug or alcohol abuse, criminal history, and record concerning appearance at court proceedings; and

        (B)whether, at the time of the current offense or arrest, the person was on probation, on parole, or on other release pending trial, sentencing, appeal, or completion of sentence for an offense under Federal, State, or local law; and

    (4)the nature and seriousness of the **danger to any person or the community** that would be posed by the person's release. In considering the conditions of release described in subsection (c)(1)(B)(xi) or (c)(1)(B)(xii) of this section, the judicial officer may upon his own motion, or shall upon the motion of the Government, conduct an inquiry into the **source of the property** to be designated for **potential forfeiture** or offered as collateral to secure a bond, and shall decline to accept the designation, or the use as collateral, of property that, because of its source, will not reasonably assure the appearance of the person as required."

"§ 3142(c)(1)(b) In any case that involves **a minor victim** under section 1201, 1591, 2241, 2242, 2244(a)(1), 2245, 2251, 2251A, 2252(a)(1), 2252(a)(2), 2252(a)(3), 2252A(a)(1), 2252A(a)(2), 2252A(a)(3), 2252A(a)(4), 2260, 2421, 2422, 2423, or 2425 of this title, or a failure to register offense under section 2250 of this title, any release order shall contain, **at a minimum**, a condition of **electronic monitoring** and each of the conditions specified at subparagraphs (iv), (v), (vi), (vii), and (viii)."

2/ Cal.Penal Code § 1320.35

**California Penal Code (Effective: January 1, 2021)**

**Pretrial risk assessment tools; legislative intent; definitions; validation; public information; report on outcomes and potential biases in pretrial release**

This statute lists the requirements for regular validation and disclosure of the data and performance of pretrial risk assessment tools. Regarding **group fairness**, § 1320.35(a) explicitly seeks to reduce group bias (unfairness) for the following sensitive features: **gender, race, or ethnicity**. In terms of disclosure, § 1320.35(d)(1)(A) requires not only the features ('line items') and risk prediction ('scoring'), but also the 'weighting' of how much each feature contributes to the prediction. Although ZIP Code of residence is not an explicit objective of bias mitigation, § 1320.35(f)(3)(B) requires model disclosure in terms of risk level aggregated by ZIP Code. It is a well known issue that even if an AI model does not use the protected feature race, the real-world bias can still penetrate into the model's prediction via the use of features that highly correlate with race, e.g. ZIP Code.[1] Therefore, this requirement suggests that California does not only care about direct bias, i.e. bias caused by the use of a sensitive feature, but also **indirect bias**, i.e. bias caused by a proxy or highly correlated feature. Regarding individual fairness, in terms of **horizontal individual fairness**, i.e. similar individuals should get similar predictions[2], § 1320.35(g)(3) does allow the release of individual-level data but under strict requirements of a research contract and anonymization. In terms of **vertical individual fairness** (vertical equity), i.e. the question of how to appropriately treat people who differ with respect to one feature on a spectrum such as income level[3], § 1320.35(f)(4) requires disclosure of "any disparate effect in the tools based on income level". In addition, § 1320.35(f)(3)(B) and § 1320.35(f)(3)(B) require the disclosure of the predictive accuracy and the risk levels aggregated by offense type. If offense type can be treated as a spectrum by level of seriousness (e.g. from infraction, misdemeanor to felony), vertical individual fairness questions can also be formulated with respect to this feature.

**Relevant excerpts from the statute:**
"§ 1320.35(a) It is the intent of the Legislature in enacting this section to **understand and reduce biases** based on **gender, income level, race, or ethnicity** in pretrial release decision making."

"§ 1320.35(d)(1) In order to increase transparency, a pretrial services agency shall, with regard to a pretrial risk assessment tool that it utilizes, make the following information publicly available:
(A) **Line items**, **scoring**, and **weighting**, as well as details on **how each line item is scored**, for each pretrial risk assessment tool that the agency uses."

---

[1] Corbett-Davies, S., & Goel, S. (2018). The measure and mismeasure of fairness: A critical review of fair machine learning. https://arxiv.org/abs/1808.00023
[2] Dwork, C., Hardt, M., Pitassi, T., Reingold, O., & Zemel, R. (2012, January). Fairness through awareness. In Proceedings of the 3rd innovations in theoretical computer science conference (pp. 214-226). https://dl.acm.org/doi/10.1145/2090236.2090255
[3] Black, E., Elzayn, H., Chouldechova, A., Goldin, J., & Ho, D. (2022, June). Algorithmic fairness and vertical equity: Income fairness with IRS tax audit models. In 2022 ACM Conference on Fairness, Accountability, and Transparency (pp. 1479-1503). https://dl.acm.org/doi/pdf/10.1145/3531146.3533204

"§ 1320.35(f) Beginning on or before June 30, 2021, and on or before June 30 of each year thereafter, the Judicial Council shall publish on its internet website a report with data related to outcomes and potential biases in pretrial release. The report shall, at a minimum, include:
(3) The following information on each risk assessment tool:
…
(B) **Risk levels aggregated** by race or ethnicity, gender, offense type, ZIP Code of residency, and release or detention decision.
(C) The **predictive accuracy** of the tool by gender, race or ethnicity, and offense type.
…
(4) If feasible, the Judicial Council shall provide information on any **disparate effect** in the tools based on **income level**."

"1320.35(g)(3) The Judicial Council shall **not share** any **individual-level data** with any outside entity unless it has entered into a contract for research purposes with the entity and privacy protections are established to anonymize the data."

3/ N.Y. Correct. Law § 168-l

**Consolidated Laws of New York - Correction Law (Effective: September 23, 2011). Board of examiners of sex offenders**

This statute lays out features to be considered when evaluating the recidivism risk of one particular category of defendants (sex offenders). One particular feature that may cause controversy is the age at the first sex offense, since a recent district court decision **Flores v. Stanford** (District Court, S.D. New York, 09/28/2021), 18 Civ. 02468 (VB)(JCM), which is in the same jurisdiction as this statute, acknowledges that the use of age of offense in recidivism risk assessment might be ground for a constitutional claim against negative parole decisions based on AI recidivism risk assessment: "**the weight COMPAS affords to youth** in determining recidivism risk will be **highly relevant** to the underlying constitutional claims".

**Relevant excerpts from the statue:**
"The board shall develop guidelines and procedures to assess the risk of a repeat offense by such sex offender and the threat posed to the public safety. Such guidelines shall be based upon, but not limited to, the following:
…
(v) the **age** of the sex offender at the **time of the commission of the first sex offense**
…"

**4/ N.J. Stat. § 2A:162-25.**

**New Jersey Statutes. Statewide Pretrial Services Program; establishment; risk assessment instrument; monitoring of eligible defendants on conditional release (Proposed Legislation)**
Related to our fairness research, this statute lists demographic features that a state-wide risk assessment tool should collect (race, ethnicity, **gender,** or socio-economic status), which might be understood as relevant for later validation study to evaluate potential 'discriminatory' impact on groups with respect to one of those features.

There are two interesting observations in this statute. Firstly, the explicit ban on discriminatory recommendation based on gender for risk assessment, if interpreted as a ban on use of 'gender' as a feature in an AI recidivism assessment tool, is at odds with some other jurisdictions. For example, **State v. Loomis** (371 Wis.2d 235, Supreme Court of Wisconsin, 10/28/2016) explicitly advocates the use of gender by COMPAS: "We determine that **COMPAS's use of gender** promotes accuracy that ultimately inures to the benefit of the justice system including defendants." Secondly, although information about financial resources of the defendants is collected, the statute does not prohibit bail discrimination based on this feature. One potential explanation is because the amount of bail should be somewhat proportional to a defendant's net worth to effectively discourage them from fleeing, e.g. a billionaire may not care about forfeiting a bail of $1 million (very high for most people) in exchange for their freedom.

**Relevant excerpts from the statue:**
"(2) The approved risk assessment instrument shall **gather demographic information** about the eligible defendant including, but not limited to, race, ethnicity, **gender,** financial resources, and socio-economic status. **Recommendations for pretrial release** shall **not be discriminatory** based on race, ethnicity, **gender,** or socio-economic status."

**5/ 725 Ill. Comp. Stat. 5/110-6.4.**

**Smith-Hurd Illinois Compiled Statutes. Statewide risk-assessment tool (Effective: January 1, 2023)**
This statute also defines risk in terms of threat to public safety (**recidivism**) or **failure to appear**. Regarding **group fairness**, this statute lays out **sensitive features** (race and gender) and non-sensitive but **highly-correlated, i.e. proxy, features** (educational level, socio-economic status, and neighborhood) that the state-wide risk assessment model cannot discriminate on. This list indicates that Illinois strives to tackle both direct and indirect biases.

**Relevant excerpts from the statue:**
"The Supreme Court shall consider establishing a risk-assessment tool that does not discriminate on the basis of **race, gender, educational level, socio-economic status,** or **neighborhood**."

6/ Ind. Code § 35-33-8

**Indiana Code. Bail and Bail Procedure (Effective: July 1, 2017)**
Section IC 35-33-8-0.5 (**Statewide evidence based risk assessment system**) lays out two risk scores that an AI model may predict to assist court officials in making bail decisions: **criminal recidivism** and **failure to appear**, and assigns the task of establishing rules on how a state-wide risk assessment tool to the the Indiana Supreme Court. Section IC 35-33-8-3.8 (Bail setting based on pretrial risk assessment system) draws a simple decision tree on conditions under which a defendant cannot be released on their own recognizance (ROR, i.e. released without bail): a **serious current charge** (murder or treason) or a temporal overlap between the current arrest and a **previous conditional release** (pretrial, probation, parole, or community supervision).

**Relevant excerpts from the statue:**
"IC 35-33-8-0.5(c) The Indiana pretrial risk assessment system shall be designed to assist the courts in assessing an arrestee's likelihood of:
(1) committing a new criminal offense; or
(2) failing to appear."

"IC 35-33-8-3.8(b) If the court finds, based on the results of the Indiana pretrial risk assessment system (if available) and other relevant factors, that an arrestee does not present a substantial risk of flight or danger to the arrestee or others, the court shall consider releasing the arrestee without money bail or surety, subject to restrictions and conditions as determined by the court, unless one (1) or more of the following apply:
(1) The arrestee is charged with murder or treason.
(2) The arrestee is on pretrial release not related to the incident that is the basis for the present arrest.
(3) The arrestee is on probation, parole, or other community supervision."

# Regulations (executive)

**1/ Ill. Admin. Code tit. 20, § 1905.60**
**Illinois Administrative Code. Adult Sex Offender Evaluation and Treatment - Standards of Practice - Risk Assessment**
Date: January 01, **2017**

Due to the specific nature of sexual offenses, specific sexual recidivism risk assessment standards have been developed, which may sometimes differ from general recidivism risk assessment standards. This regulation lays out the standards in Illinois for human professionals to conduct recidivism risk assessments of sexual abusers. In particular, the regulation groups relevant features into six categories: 1. Criminal history, 2. Victim-related variables, 3. Sexual deviancy, 4. Antisocial orientation, 5. Intimacy and relationship deficits, and 6. Self-regulation difficulties. Interestingly, the regulation requires professionals to use "validated actuarial risk assessment tools" and "structured, empirically guided risk assessment protocols" together, which may imply that if features suggested to go into the mental models of human professionals can be effectively represented by numerical values, such features should also be used as input features to an AI sexual recidivism risk assessment tool. For example, while quantitative categories such as Criminal history category (e.g. number of prior arrests or convictions) and Victim-related variables (e.g. age, gender, relationship) can be used for the AI model, variables in the other four qualitative categories might first be processed by the professionals to get intermediate quantitative scores for each of those variables, which might then be fed into an AI model as additional input data indicative for sexual recidivism prediction.

**Relevant excerpts from the regulation:**
"...
b) Evaluators conducting risk assessments on sexual abusers are well versed in the contemporary research regarding static and dynamic factors linked to recidivism among sexual abusers. These variables fall into the following categories:
1) Criminal history (e.g., prior arrests, convictions);
2) Victim-related variables (e.g., age, gender, relationship);
3) Sexual deviancy (e.g., offense-related sexual arousal, interests and/or preferences; sexual preoccupation);
4) Antisocial orientation (e.g., criminal attitudes, values and behaviors; lifestyle instability);
5) Intimacy and relationship deficits (e.g., problems with intimacy, unstable relationships, conflictual intimate relationships, deficits in social support and interaction); and
6) Self-regulation difficulties (e.g., hostility, substance abuse, impulsivity, access to victims).

c) Evaluators conducting risk assessments of sexual abusers use empirically supported instruments and methods (i.e., validated actuarial risk assessment tools and structured, empirically guided risk assessment protocols) over unstructured clinical judgment.
…"

Pennsylvania Administrative Code
**Sentence Risk Assessment Instrument methodology**
Date: Dec. 20, **2019**

This regulation describes the technical procedure carried out in Pennsylvania to select which features are the most useful to predict recidivism and what those most predictive features are. The technical procedure includes the following steps: bivariate analyses, multivariate logistic regression (only at the initial features' predictive power estimation phase, not for the final risk assessment model), categories rotation, analysis of the ROC (true positive vs. false positive) curve, and final scale validation. The highly predictive features reported at the end are age, gender, number of prior convictions, offense types of prior and current convictions, multiple current convictions, and prior juvenile adjudication. This list does not include race (a 'strict scrutiny' feature) but does include gender and age (an 'intermediate scrutiny' and a 'rational basis' feature, respectively), which may contribute to the discussion that although there exists an implicit consensus among AI recidivism risk assessment tool developers and the public that race should not be used as a feature, some jurisdictions (e.g. Pennsylvania as illustrated in this regulation) advocate for the use of age and gender (or more generally, features that do not belong to the 'strict scrutiny' category) while others may not.

Interestingly, though not directly related to the fairness question, the risk scales in § 305.7, which is referred to at the end of § 305.2 as the resulting model from this methodology regulation, is a rather simplistic but more interpretable model (decision tree, e.g. 'male' gives an overall risk score increment of 1 while 'female' gives an increment of 0) compared to most AI recidivism models (e.g. end-to-end logistic regression models). This explicit use of a decision tree in this regulation may open room for discussions of the model complexity vs. interpretability tradeoff, which can be linked to the fairness question because model complexity often correlates with accuracy, and certain group fairness metrics strive for equalized accuracy across groups.


**Relevant excerpts from the regulation:**
"…
(3) In developing the risk scales, the following analyses were conducted:
(i) bivariate analyses to determine which factors were related to recidivism;
(ii) multivariate logistic regression to determine which factors best predicted recidivism while holding other factors constant;
(iii) rotation of all categories for factors that were multi-categorical to ensure that reported differences were real and not due to a particular comparison category;
(iv) Receiver Operating Characteristic (ROC) analysis, which plots the true positive rate (i.e., how many people were predicted to recidivate and did recidivate) against the false positive rate (i.e., how many people were predicted to recidivate but did not recidivate); and
(v) validation of the final scales with both samples.
(b) *Risk factors and scales--general*.
(1) Based upon the analyses conducted by the Commission, the following factors were found to be predictive of recidivism, and thus, used in the risk assessment scales:

(i) age;
(ii) gender;
(iii) number of prior convictions;
(iv) prior conviction offense type;
(v) current conviction offense type;
(vi) multiple current convictions;
(vii) prior juvenile adjudication.
(2) The risk scale for recidivism is located at § 305.7."

3/ Or. Admin. R. 291-078-0020
Oregon Administrative Code
**Case Management System (Community Corrections) - Risk Assessment**
Date: Feb 25, **2015**

This regulation specifies the recidivism risk assessment tools and procedure used on offenders who are to be released on probation (under community supervision) in Oregon. What interesting about this regulation compared to similar recidivism risk assessment regulations from other states is that they require the assessment to involve more than one tool, e.g. for general offenses, one initial tool (such as PSC or Proxy) with only three risk level outputs: low, medium, high) and a more thorough tool (such as LS/CMI) to be used if the risk level returned by the initial tool is at least medium. The more thorough tool (LS/CMI) takes into account a variety of factors which are grouped into eight domains: Criminal History, Education/Employment, Family/Marital, Leisure/Recreation, Companions, Alcohol/Drug Problem, Procriminal Attitude/Orientation, Antisocial Pattern. The two-phase risk assessment procedure reduces the risk of bias from one model to affect the final risk outcome of a defendant from a minority group, provided that the tool at each phase is developed independently (e.g. in terms of training and model selection). However, one caution is that if the source of unfairness comes from the historically biased training data of Oregon past inmates, the use of two different tools may not address this issue. Another interesting observation from this regulation is that only medium or high risk offenders identified by the first tool are required to be assessed by the second tool, which indicates that the Oregon government cares much more about minimizing false positives (i.e. lower-risk defendants misclassified by the two-phase pipeline as higher-risk) that they might tolerate some false negatives. This regulation is therefore consistent with the common law principle of criminal justice systems as concisely put by William Blackstone in 1769: "the law holds that it is better that 10 guilty persons escape, than that 1 innocent suffer" or the high standard of proof in criminal proceedings (e.g. "beyond reasonable doubt").

**Relevant excerpts from the regulation:**
"(1) Proper assessment ensures the classification of offenders according to risk and their assignment to specified levels of community supervision. The following risk assessment tools are utilized by the department and county community corrections agencies for risk assessment of offenders:
(a) Level of Service/Case Management Inventory (LS/CMI) Section 1 General Risk/Need Factors (version Feb. 2013): A validated assessment tool used to determine an offender's risk to recidivate and identify criminogenic risk factors across eight domains Criminal History, Education/Employment, Family/Marital, Leisure/Recreation, Companions, Alcohol/Drug Problem, Procriminal Attitude/Orientation, Antisocial Pattern.
…
(c) Public Safety Checklist (PSC)(version 2005): A statistical calculation developed by the Oregon Criminal Justice Commission in collaboration with the department's research unit to predict an offender's risk to recidivate within three years of release from custody or admission to probation.
(d) Proxy (version 2005): A three question validated risk assessment tool used to identify initial risk for offenders entering probation supervision.
…

(4) Risk, Needs, and Responsivity Assessment:

(a) The ongoing assessment of offenders risk, needs, and responsivity relies on a combination of both static and dynamic risk factors in order to predict recidivism and identify criminogenic needs and responsivity issues.

(b) The LS/CMI and a case plan, as described in OAR 291-078-0026, will be completed on all offenders determined to be of high or medium risk either by the PSC, Proxy, or by an approved override. The LS/CMI is not required on sexual offenders who are subject to the Stable/Acute and Static-99R.

…"

**4/ Cal. Code Regs. tit. 15, § 3768.1**
California Code of Regulations
**California Static Risk Assessment**
Date: January 07, **2010**, including amendment of subsection (b)(5) filed on July 22, 2010 (Register 2010, No. 30)

This regulation lays out the main criteria and outputs for the California Static Risk Assessment (CSRA) tool in the context of felony recidivism prediction in the three years following a parole. The regulation first identifies a non-exhaustive list of main features to be used by the tool: age, gender, and criminal history ("criminal misdemeanor and felony convictions, and sentence/supervision violations"). One unique aspect of this regulation is that even though the low risk and moderate risk outputs returned by CSRA are generic labels as in most other recidivism risk assessment tools, there is no single high risk label, but rather three categories: "High Risk Drug", "High Risk Property", and "High Risk Violence". This innovation may make the AI model more interpretable and thus facilitate the validation process for AI experts to notice when the model makes critical mistakes, e.g. giving an exceptionally high weight to the number of DUI offenses in a person's criminal history to justify a "High Risk Property" prediction. Furthermore, the high risk label is the most likely to mandate court officials to either reject the parole application or grant the parole but with stringent government and community supervision protocols. Specifying the type of offense a defendant has high risk to commit will firstly better equip judicial officials to decide the appropriate type of additional parole supervision, e.g. mandatory regular visits to rehabilitation facilities for inmates classified as "High Risk Drug", and secondly to identify which feature is or is not relevant to evaluate group fairness metrics with respect to. For example, gender is often not relevant to evaluate group fairness metrics with respect to the "High Risk Violence" label because women are often less prone to violence than men and most people do not expect an equalized positive output rate of "High Risk Violence" across male and female inmates.

**Relevant excerpts from the regulation:**
"(a) The California Static Risk Assessment (CSRA) (new 12/09 ), which is incorporated by reference, is a validated risk assessment tool that utilizes a set of risk factors which are most predictive of recidivism. The tool produces a risk number value that will predict the likelihood that an offender will incur a felony arrest within a three-year period after release to parole.
Risk factors utilized include, but are not limited to, age, gender, criminal misdemeanor and felony convictions, and sentence/supervision violations.
(b) CSRA risk number values fall in one of following five categories:
(1) Low Risk, with a risk number value of "1".
(2) Moderate Risk, with a risk number value of "2".
(3) High Risk Drug, with a risk number value of "3". High Risk Drug means that the offender has a greater risk of reoffending with a drug offense.
(4) High Risk Property, with a risk number value of "4". High Risk Property means that the offender has a greater risk of reoffending with a property offense.
(5) High Risk Violence, with a risk number value of "5". High Risk Violence means that the offender has a greater risk of reoffending with a violent offense.
…"

5/ La. Admin. Code Pt V, 7525
Louisiana Administrative Code
**Juvenile Detention Facilities - Data**
Date: July **2012**

This regulation is related to a more specific sub-population of defendants in Louisiana: juvenile offenders. It lists the information (features) of juvenile offenders that the detention facilities must collect, which includes admission data (Sub-section A), operational data (Sub-section B), and detention screening data (Sub-section C). Although the regulation does not specify which features to be inputs of an AI risk assessment model, Sub-section C lists features that must be collected if a Risk Assessment Instrument (RAI) is used (e.g. demographic features such as race, ethnicity, gender, date of birth, with which age can be computed, residence, offense type, screen data, and outcome in terms of any recidivism or failure to appear) from which we may decide AI-usable features and/or group fairness vs. individual fairness evaluation features. In particular, a group fairness evaluation feature may help us consider whether a group fairness metric is equalized across groups of that feature, e.g. whether the 'detention' recommendation rate is equalized across black and white juveniles. On the other hand, an individual fairness evaluation feature may help us construct a distance function to quantify how similar two individuals are (with respect to features we deem relevant for the output), which is a crucial first step for the individual fairness principle that similar individuals should get the same treatment (output).

**Relevant excerpts from the regulation:**
"...
C. Detention Screening Data

1. If a provider conducts a Risk Assessment Instrument (RAI) on new admissions, it shall maintain an accurate record of the following data fields:
a. demographics of youth screened, aggregated by:
i. race;
ii. ethnicity;
iii. gender;
iv. date of birth;
v. parish of residence; and
vi. geographical zone determined by provider to include zip code, local law enforcement zones;
b. offense of youth screened:
i. specific charge(s); and
ii. release date;
c. screen data:
i. date completed;
ii. overrides usage; and
iii. screening outcomes: release/alternative to detention/secure detention;
d. outcome data:
i. successful/unsuccessful; and
ii. recidivism/failure to appear (FTA)."

**6/ Or. Admin. R. 255-085-0020**
Oregon Administrative Code
**Sex Offender Risk Assessment Methodology**
Date: August 27, 2015 (last amended on August 16, 2022)

This regulation shows a map from risk scores returned by an AI model (Static-99R) for sex offenders to the level of notification such offenders are required to maintain with the government as part of their sex offender registration process after having served their sentence. An intriguing fairness angle from this regulation is that with respect to gender, due to the specific nature of sexual offenses, different standards of risk assessment are applied to male vs. female. For instance, prostitution-related arrests, charges, or convictions count as an "evidence-based risk factor" that disqualify a female defendant from being assigned to the Notification Level 1 (lowest recidivism risk). However, this example raises the question of whether such gender-specific standards, which may disadvantage women, are based on legitimate biological differences or unfounded social stigma against one gender.

**Relevant excerpts from the regulation:**
"(1) Classifying agencies shall place each registrant into one of the following levels:
(a) Notification Level 1: A registrant who presents the lowest risk of reoffending and requires a limited range of notification;
(b) Notification Level 2: A registrant who presents a moderate risk of reoffending and requires a moderate range of notification; or
(c) Notification Level 3: A registrant who presents the highest risk of reoffending and requires the widest range of notification.

(2) For classification and community notification for adult male registrants, classifying agencies shall use the Static-99R actuarial instrument with the coding manual, Exhibit STATIC-99R, to conduct a sex offender risk assessment, except as to where it conflicts with OAR 255-085-0020(6). Classifying agencies may score registrants using information from previous Static-99 or Static-99R assessments. Classifying agencies shall score and place each registrant into a notification level:
(a) Notification Level 1: Static-99R score of -3 to 3;
(b) Notification Level 2: Static-99R score of 4 to 5; or
(c) Notification Level 3: Static-99R score of 6 or higher.
…

(4) Level 1 Classification for Certain Registrants:
…
(b) Classifying agencies may classify a **female** registrant into Level 1, without using the methodologies listed in OAR 255-085-0020 (2) or (3) or (4)(a), unless evidence-based risk factors exist to indicate that the female registrant is at a higher risk to reoffend sexually and a higher level of notification may be appropriate. Evidence-based risk factors for sexually reoffending for a female registrant may include:
(A) The registrant has an arrest, charge, or conviction for a child abuse offense;
(B) The registrant has an arrest, charge, or conviction for promoting prostitution or compelling prostitution;

(C) The registrant has a conviction for a person felony or Class A person misdemeanor as defined by the rules of the Oregon Criminal Justice Commission, subsequent to the registrant's initial conviction that required sex offender registration;

(D) The registrant has an arrest, charge, or conviction for a crime that would require registration as a sex offender in addition to the registrant's conviction that required sex offender registration; or

(E) The registrant has repeated (2 or more) criminal convictions for any offense resulting from separate criminal episodes in the five years preceding the classification.…"



\end{document}